\documentclass[reprint,amsmath,amssymb,aps,nofootinbib,noeprint]{revtex4-2}

\usepackage{graphicx}
\usepackage{epstopdf}
\usepackage{aas_macros}
\usepackage{amsmath}
\usepackage{xcolor}
\usepackage{ragged2e}
\usepackage{MnSymbol}
\usepackage{mathrsfs}

\usepackage{multirow}
\usepackage{booktabs}

\definecolor{darkgreen}{rgb}{0.0,0.5,0.0}
\usepackage[colorlinks,citecolor=darkgreen]{hyperref}

\usepackage[leftmargin=0pt,rightmargin=0pt,skipabove=2pt,skipbelow=2pt,innerleftmargin=5pt,innerrightmargin=5pt,innertopmargin=5pt,innerbottommargin=5pt,linewidth=0pt,innerlinewidth=0pt,middlelinewidth=0pt,outerlinewidth=0pt,roundcorner=0pt]{mdframed}

\newcommand{\ie}{\emph{i.e.} }
\newcommand{\eg}{\emph{e.g.,} }

\newcommand{\be}{\begin{equation}}
\newcommand{\ee}{\end{equation}}
\newcommand{\bea}{\begin{equation*}}
\newcommand{\eea}{\end{equation*}}
\newcommand{\beqr}{\begin{eqnarray} \nonumber}
\newcommand{\eeqr}{\end{eqnarray}}
\newcommand{\beqrb}{\begin{eqnarray}}
\newcommand{\eeqrb}{\nonumber \end{eqnarray}}
\newcommand{\fin}{\mbox{ .}}
\newcommand{\coma}{\mbox{ ,}}

\newcommand{\cm}{\mbox{ cm}}
\newcommand{\sr}{\mbox{ sr}}
\newcommand{\se}{\mbox{ s}}

\newcommand{\erg}{\mbox{ erg}}

\newcommand{\keV}{\mbox{ keV}}

\newcommand{\GeV}{\mbox{ GeV}}
\newcommand{\TeV}{\mbox{ TeV}}

\newcommand{\const}{\mbox{const}}

\newcommand{\gama}{$\gamma$}

\newcommand{\mynewcommand}[2]{\ifdefined #1 \else \newcommand{#1}{#2} \fi}

\mynewcommand{\apj}{ApJ}     
\mynewcommand{\apjl}{ApJL}     
\mynewcommand{\apjs}{ApJS}    
\mynewcommand{\aap}{A\&A}    
\mynewcommand{\nat}{Nature}  

\newcommand{\dgr}{^{\circ}}

\newcommand{\dgrdot}{{\overset{^\circ}{.}}}
\newcommand{\mindot}{{\overset{{'}}{.}}}

\setcitestyle{super}

\makeatletter
\def\@hangfrom@section#1#2#3{\@hangfrom{#1#2}#3}
\def\@hangfroms@section#1#2{#1#2}
\makeatother

\newcommand{\TblNote}[1]{\par\smallskip{\footnotesize\justifying\noindent #1\par}}

\begin{document}
\raggedbottom

\title{\emph{Fermi}--eROSITA cross-correlations suggest annihilation lines of $\sim70$ GeV WIMPs}

\author{Uri Keshet}
\address{
    Physics Department, Ben-Gurion University of the Negev, POB 653, Be'er-Sheva 84105, Israel
}
\email{keshet.uri@gmail.com}

\begin{abstract}
\begin{mdframed}[backgroundcolor=black!5]
The deep potential wells of galaxy clusters provide some of the strongest constraints on the velocity-dependent ($p$-wave) annihilation of weakly interacting massive particle (WIMP) dark matter.
Cross-correlating high-latitude \emph{Fermi}-LAT 16.3-year data with the western Galactic hemisphere eROSITA map facilitates high-sensitivity tests.
A blind composite matched-filter scan of the cross-correlated spectrum for the $\chi\chi\to\gamma\gamma$, $\gamma Z$, and $\gamma h$ annihilation-line triad of a single WIMP $\chi$ finds $m_\chi c^2\simeq71.9$ and $67.5$ GeV (observer-frame) triads, each with a local $Z$-score $\sim3$ and a (henceforth global) $1.7$--$2.1\sigma$ (trial-corrected), insensitive to details of detrending, sky masks, and filters.
The leading nonad (equidistant three triads) corresponds to $66.4$ and $71.9$ GeV cross-annihilating WIMPs ($2.5\sigma$).
The leading $\sim72$ GeV triad coincides with the mass implied by interpreting the reported $\sim43.2$ GeV $\gamma$-ray line in three nearby clusters as its $\gamma Z$ channel.
At this externally fixed mass, the triad reaches $3.0\sigma$ alone, and $5.1\sigma$ when combined with the reported line ($3.8\sigma$ under alternative assumptions; similar values for the nonad).
This triad, absent from the ambient (X-ray-uncorrelated) spectrum, presents in different Galactic sectors, strengthens under cosmological (positive) but not negative redshift broadening $\Delta z$, and peaks at $\Delta z\simeq 0.05$, consistent with galaxy clusters; its two non-$43$ GeV channels reach $2.4\sigma$ by themselves.
Measured (not loss-corrected) intrinsic cross sections are a few $10^{-19}$ cm$^3$ s$^{-1}$ for WIMPs tracing the X-ray gas.
\end{mdframed}
\end{abstract}

\maketitle

\section{Introduction}
\label{sec:Intro}

Evidence for dark matter (DM) first emerged from the kinematics of the solar neighborhood \cite{Kapteyn1922, Oort1932}, galaxy rotation curves \cite{Babcock1939, RubinFord70}, and the virialized motions of galaxy clusters \cite{Zwicky33}, later corroborated by gravitational lensing by massive structures \cite{TysonEtAl90, CloweEtAl06} and X-ray observations of intracluster gas \cite{FabricantEtAl80}.
Cosmological evidence requires non-baryonic cold DM to explain the rapid growth of large-scale structure (LSS) \cite{BlumenthalEtAl84}, temperature anisotropies in the cosmic microwave background (CMB) \cite{Peebles1982, SmootEtAl92}, and big bang nucleosynthesis (BBN) baryon density limits \cite{GottEtAl74}.
X-ray imaging of cluster collisions strikingly demonstrated the collisionless nature of DM, both by comparison with lensing mass reconstructions \cite{CloweEtAl06} and on purely kinematic grounds \cite{KeshetEtAl21}.

Leading theoretical DM candidates\cite{BertoneHooper18, ArbeyMahmoudi21} include weakly interacting massive particles (WIMPs) \cite{LeeWeinberg77}, massive astrophysical compact halo objects (MACHOs) \cite{AlcockEtAl00MACHO}, axions \cite{PreskillEtAl83}, primordial black holes (PBHs) \cite{Chapline1975, CarrEtAl21PBH}, sterile neutrinos \cite{DodelsonWidrow94}, and ultra-light fuzzy DM \cite{HuEtAl00}.

WIMP searches, in particular, proceed across multiple complementary frontiers. Direct detection experiments search for DM scattering via nuclear and electron recoil \cite{AalbersEtAl23LZ, AprileEtAl23XENONnT, AgnesEtAl18DarkSide, AgneseEtAl18SuperCDMS}, while accelerator-based experiments
hunt for missing momentum, invisible decay, and beam-dump scattering signatures \cite{AadEtAl21ATLAS, SirunyanEtAl21CMS, LeesEtAl17BaBar, BanerjeeEtAl19NA64, AguilarArevaloEtAl17MiniBooNE}.
Canonical astrophysical limits arise from \emph{Fermi}-LAT (henceforth LAT) observations of dwarf spheroidals (dSphs), our Galactic center (GC), massive galaxy clusters, and the isotropic \gama-ray background \cite{Ackermann2015_FermiLAT, AlbertEtAl17, AckermannEtAl17GC, AckermannEtAl15Clusters, AckermannEtAl15IGRB}, imaging atmospheric \smash{\v{C}erenkov} telescopes pointed at the GC, satellite galaxies, and local clusters \cite{Abdalla2022_HESS, AhnenEtAl16, ArchambaultEtAl17VERITAS, AbramowskiEtAl12HESS}, neutrino detectors \cite{AartsenEtAl17IceCube, AlbertEtAl17ANTARES, ChoiEtAl15SuperK}, and cosmic-ray measurements \cite{ChangEtAl08ATIC, AdrianiEtAl09PAMELA, AguilarEtAl16AMS}.

WIMPs are confined to the $1\GeV \lesssim m_\chi c^2 \lesssim 100\TeV$ mass range by early-Universe constraints and perturbative unitarity if they are thermal relics \cite{LeeWeinberg77, GriestKamionkowski90}; otherwise the main restriction is $m_\chi c^2\gtrsim 5\keV$, for DM cold enough to drive LSS formation \cite{BlumenthalEtAl82, IrsicEtAl17}.
Here, WIMPs are denoted $\chi$, and $c$ is the speed of light.

CMB observations severely constrain velocity-independent ($s$-wave) annihilation in the early Universe, imposing a $\overline{\sigma v}_s \lesssim 10^{-27.5}[m_\chi c^2/(f_{\text{eff}}\mbox{ GeV})]\text{ cm}^3\text{ s}^{-1}$ upper limit on any WIMP \cite{AghanimEtAl20Planck}.
Here, $v$ is the relative velocity, $\sigma$ is the cross section, $f_{\text{eff}}\simeq0.1$--$0.6$ (depending on the channel; lower if invisible channels dominate) is the evolution-weighted fraction of annihilated $2m_\chi c^2$ energy deposited in the plasma near redshift $z\simeq600$, and the overline on $\overline{\sigma v}$ represents velocity averaging.
At $10\lesssim m_\chi c^2/\mbox{GeV}\lesssim 100$ masses, dSphs impose a stronger $\sim 10^{-26}\cm^3\se^{-1}$ upper limit \cite{Ackermann2015_FermiLAT, AlbertEtAl17, HoofEtAl20dSphs}, comparable to the continuum cross section suggested for the GC excess \cite{CaloreEtAl15, AlemannoEtAl26}.
Such $s$-wave constraints have motivated alternative frameworks featuring suppressed or entirely forbidden $s$-wave annihilation \cite{GriestSeckel91, KumarMarfatia13, BerlinEtAl14, DAgnoloRuderman15}.

The next leading-order, velocity-dependent ($p$-wave) intrinsic cross section $\overline{\sigma v}_p$ corresponds to an annihilation rate proportional to an effective cross section \cite{DiamantiEtAl14, AbdullahEtAl14}
\begin{equation}\label{eq:EffCS}
\overline{\sigma v}_{\text{eff}} = \overline{\sigma v}_p \,\beta^2 \,B \, ,
\end{equation}
where $\beta$ is the dimensionless line-of-sight velocity dispersion of DM in the halo, and $B$ is a DM boost factor (\eg due to substructure).
For substructure-free halos ($B=1$, used for normalization henceforth), the $\beta^2 \sim 10^{-9}$ dSph limits translate\cite{ZhaoEtAl18pWave} to $\overline{\sigma v}_p\lesssim10^{-16}\cm^3\se^{-1}$ at $m_\chi c^2\simeq70\GeV$.
In the $\beta^2\sim10^{-7}$ GC, the GeV excess corresponds to $\overline{\sigma v}_p\sim10^{-19}\cm^3\se^{-1}$ for a DM cusp \cite{CaloreEtAl15}, but
higher values are obtained for alternative models of the excess \cite{KarwinEtAl17} or for a DM core.

Strong limits are imposed by $\beta^2 \sim 10^{-5}$ galaxy clusters. Their $s$-wave limits \cite{AbramowskiEtAl12HESS, AleksicEtAl10MAGIC, VERITAS12_Coma} correspond, by Eq.~\eqref{eq:EffCS}, to $\overline{\sigma v}_p\sim10^{-18}\cm^3\se^{-1}$.
Even stronger limits down to $\sim10^{-19}\cm^3\se^{-1}$ follow from LAT cluster analyses \cite{AckermannEtAl15Clusters}, but only assuming extended cluster emission following a prescribed spatial profile, and using global diffuse background templates which may over-subtract the signal\cite{KeshetEtAl04_EGRB, CaloreEtAl15}.

This $\overline{\sigma v}_{p}$ regime, $\sim6$ orders of magnitude above the canonical thermal relic ($\beta^2\simeq 0.05$) cross-section requirement \cite{LeeWeinberg77, ScherrerTurner86}, is motivated by WIMP models with enhanced late-Universe annihilation, \eg via Sommerfeld effects \cite{HisanoEtAl04, ArkaniHamedEtAl09}, Breit--Wigner resonances \cite{GriestSeckel91, IbeEtAl09}, bound-state formation \cite{PospelovRitz09}, and large $B$ factors \cite{Lake90, SilkStebbins93}, or for a relic abundance unregulated by thermal freeze-out, \eg in non-standard cosmological histories \cite{KamionkowskiTurner90, GelminiGondolo06}, non-thermal production \cite{MoroiRandall20}, or asymmetric DM \cite{Nussinov85, KaplanEtAl09}.
Such models can evade the direct detection constraints
\cite{BaerEtAl15, ArcadiEtAl18}.

The $p$-wave annihilation signal is dominated by galaxy clusters and the LSS around them.
Such weak signals can be picked up via cross-correlations with LSS-tracing X-ray maps that cover large fractions of the sky, although the parameters of the radiating clusters cannot be selected.
Here we cross-correlate the LAT data with the available high-resolution eROSITA map of the western Galactic hemisphere, at high native spectral LAT resolution, and search for the kinematically constrained line triad expected from annihilating WIMPs.

Monochromatic $\gamma$-ray lines from WIMP annihilation, in particular the $\chi\chi\to\gamma\gamma$, $\gamma Z$, and $\gamma h$ triad of channels, can provide a smoking-gun signature \cite{BergstromSnellman88, JacksonEtAl10}.
No significant line has been found in LAT and DAMPE searches \cite{AckermannEtAl13, AckermannEtAl15Lines, AndersonEtAl16, AlemannoEtAl22Lines}; the $\sim130\GeV$ GC candidate \cite{Weniger12} faded in reprocessed LAT data, and a similar feature in Earth-limb data, expected to be featureless, pointed to an instrumental origin \cite{AckermannEtAl13, AckermannEtAl15Lines}.

A line-like $\sim43\GeV$ excess in nearby galaxy clusters has been noticed over the last decade \cite{LiangEtAl16, ShenEtAl21}.
Recently, a $\sim43.2\GeV$ line (denoted $L_{43}$) was reported in a stacking analysis of the Virgo, Fornax, and Ophiuchus clusters \cite{FanEtAl26}, at a global significance of $4.3\sigma$ (or $3.7\sigma$ when including ten additional nearby clusters).
An independent analysis \cite{Profumo26} reproduces the excess, but, under alternative assumptions, finds a lower, $\sim2.7\sigma$ global significance (denoted $L_{43b}$) conditional on the choice of the three clusters,
as well as a spatially broad emission, argued to disfavor a DM origin.
Note that broad emission is less problematic for $p$-wave annihilation, whose $\propto \beta^2\rho_{\chi}^2$ emissivity (for a WIMP mass density $\rho_\chi$) is expected to be more extended than the canonical $\propto\rho_{\chi}^2$.

If interpreted as a $\gamma Z$ or $\gamma h$ WIMP annihilation channel, $L_{43}$ corresponds respectively to a mean mass $\overline{m}_\chi c^2\simeq 72.1\GeV$ (nominal triad mass) or $87.8\GeV$ of two annihilating WIMPs (possibly of different species),
and implies companion triad lines at predetermined energies.
We discuss this possibility in \S\ref{sec:43GeV}, after outlining the methodology (\S\ref{sec:Methodology}) and results (\S\ref{sec:Results}).
Additional support for this interpretation is outlined in the discussion \S\ref{sec:Summary}, along with its implications, in particular for the GC GeV excess.

\section{Methodology}
\label{sec:Methodology}

High angular and energy resolution LAT \gama-ray data (\S\ref{subsec:FermiLAT}) are cross-correlated with an eROSITA X-ray map as an LSS tracer (\S\ref{subsec:Correlations}),
enhancing the sensitivity for galaxy-cluster spectral features and for DM detection.
Contaminations are minimized by an aggressive masking of all known cataloged\cite{AbdollahiEtAl20, FGL4DR4} \gama-ray sources, the Galactic plane, the inner Galaxy, and bright X-ray sources.

The resulting X-ray-correlated energy spectrum, after its smooth continuum is removed (detrended), is scanned by sliding spectral apertures (simple and composite matched filters; \S\ref{subsec:MatchedFilters}), which incorporate the asymmetric, non-Gaussian LAT energy dispersion, in search of annihilation lines as well as kinematically constrained line complexes.

Global significance, which accounts for multiple trials (the look-elsewhere effect) of scanning many energies or masses, is calibrated by randomly shuffling the X-ray contrast values among the analyzed pixels and rerunning the full pipeline, including the detrending; shuffling keeps the \gama-ray data unchanged but destroys any true correlation between the \gama-ray and X-ray maps.
Monte Carlo simulations of Gaussian noise, with the binning and empirical errors of the measured spectrum, provide an independent calibration.

Finally, line emission is converted into an annihilation cross section by approximating the velocity-weighted DM distribution as tracing the X-ray emitting gas (\S\ref{subsec:SigmaCorrelations}).
This conversion accounts for the known gas properties (such as the X-ray cooling function) and for the expected photon multiplicity $Y_i$ of each annihilation channel within a line complex, consolidating the unknown astrophysical and DM-sector variables into an ignorance factor $D$.
Biases of the inferred energies and masses, and the fractions of the line signal
lost to the point spread function (PSF), the pipeline, and X-ray noise, are estimated in \S\ref{subsec:Losses}.

\subsection{\emph{Fermi}-LAT data reduction}
\label{subsec:FermiLAT}

Pre-generated archival Pass 8 (P8R3)\cite{BruelEtAl18Pass8R3} data are extracted from the Fermi Science Support Center (FSSC)\footnote{\url{http://fermi.gsfc.nasa.gov/ssc}},
and reduced using the Fermi Science Tools (version \texttt{2.2.0}).
Weekly all-sky photon event files spanning $\sim 16.3$ years (mission weeks $9$ through $859$) are selected for the highest purity \texttt{ULTRACLEANVETO} event class (\texttt{evclass=1024}).
Corresponding exposure maps are computed using the \texttt{P8R3\_ULTRACLEANVETO\_V3} instrument response functions (IRFs).

We adopt 256 logarithmically spaced bins spanning the $1$--$200\GeV$ photon energy range, before restricting to the narrower $3$--$100\GeV$ window relevant to our WIMP annihilation-line search.
Each energy bin, of width $\Delta\epsilon$, thus spans a fractional $s_0\equiv\Delta\epsilon/\epsilon\simeq 2.1\%$ width, oversampling by a factor of $\sim3$ the LAT $68\%$ energy containment, which is nearly uniformly $\sim7\%$ in the energy range of interest\cite{AjelloEtAl21}, and by a factor of $\sim1.8$ the $\sim3.7\%$ dispersion of a Gaussian with the same, $\sim9\%$ full width at half maximum as the IRF core at $70\GeV$.
Different choices of resolution and binning are also explored
(\S\ref{subsec:Robustness} and Appendix~\ref{app:Nominal}).

This $3$--$100\GeV$ search window is primarily dictated by the high energy-resolution needed for identifying sharp spectral features.
The low limit is also required to ensure a sufficiently compact PSF.
The upper limit is needed also to keep the extragalactic background attenuation \cite{FranceschiniEtAl08, DominguezEtAl11} weak ($\sim15$--$20\%$ for $100\GeV$ photons arriving from $z=0.5$) and the photon statistics sufficient.

A maximum zenith angle cut of $90\dgr$ is applied to mitigate Earth-limb contamination from cosmic-ray interactions in the upper atmosphere\cite{AbdoEtAl09EarthLimb}. Good time intervals are defined using the standard filter expression \texttt{(DATA\_QUAL > 0 \&\& LAT\_CONFIG==1)}. The filtered events are mapped onto an order 10 ($N_{\mathrm{side}} = 1024$) full-sky HEALPix\citep{GorskiEtAl05} grid of $\sim 3\mindot4$ pixel separation and $\delta\Omega\simeq1.0\times10^{-6}\sr$ pixel solid angle.

\begin{table}[t]
    \centering
    \caption{Analysis parameters.
    \label{tab:CorrParameters}}
    \setlength{\tabcolsep}{4pt}
    \begin{tabular}{@{}l l l@{}}
        \toprule
        Property & Symbol & Value \\
        \midrule
        LAT data span & --- & $16.3$ yr \\
        Latitude cut & $|b|_{\min}$ & $30\dgr$ \\
        Longitude cut & $|l|_{\min}$ & $50\dgr$ \\
        Source mask radius & $\theta_{\min}$ & $0\dgrdot5$ \\
        HEALPix order & --- & $10$ \\
        Pixel solid angle & $\delta\Omega$ & $1.0\times10^{-6}\sr$ \\
        ROI pixels & $H$ & $1.99\times10^{6}$ \\
        Native energy bins & $s_0$ & $2.1\%$ \\
        Search window & --- & $3$--$100\GeV$ \\
        X-ray energy band & --- & $0.2$--$2.3\keV$ \\
        X-ray brightness & $\langle \boldsymbol{X} \rangle$ & $9.0\times10^{-8}\erg \se^{-1}\cm^{-2}\sr^{-1}$ \\
        Contrast filter & $\theta$ & $2\dgr$ \\
        Contrast variance & $\llangle \boldsymbol{X}_\theta^2\rrangle$ & $0.19^\dagger$ \\
        Detrending scale & $\lambda$ & $0.62$ \\
        Line window & $f$ & $0.1$ \\
        Triad window & $f$ & $0.015$ \\
        Nonad window & $f$ & $0.05$ \\
        Nonad separation & $f_t$ & $0.02$ \\
        Line scan step & --- & $0.2\%$ \\
        Triad scan step & --- & $0.1\%$ \\
        Nonad scan step & --- & $0.5\%$ \\
        Cooling function & $\Lambda(0)$ & $8.5\times10^{-24}\erg\cm^3\se^{-1}$ \\
        \bottomrule
    \end{tabular}
        \TblNote{$^\dagger$Median over energy bins.}
\end{table}

Wide-field spectral LAT analyses require an aggressive masking of \gama-ray point sources, which contribute a considerable part of the high-energy sky.
Otherwise, bright sources can imprint line-like features on the analyzed spectrum; such features have been reported in the spectra of individual active galactic nuclei (AGN) \cite{ZhouEtAl21, KangEtAl26a, Kang26b, KangEtAl26c}, and weak line-like excesses are common in blind scans of the sky \cite{LiEtAl19}.
Thus, all HEALPix pixels falling within a $\theta_{\min}=0\dgrdot5$ radius --- the $68\%$ PSF containment at $\sim2\GeV$ --- of any source in the 14-year LAT fourth source catalog (4FGL-DR4)\cite{AbdollahiEtAl20, FGL4DR4} are masked prior to the analysis.
The Galactic plane is similarly masked at latitudes $|b|<|b|_{\min}=30\dgr$, and the inner Galaxy, with its prominent Fermi and eROSITA bubbles, is avoided by masking pixels with $|180\dgr-[(l+180\dgr)\mbox{ mod }360\dgr]|<|l|_{\min}=50\dgr$.

\subsection{Cross-correlations with X-ray maps}
\label{subsec:Correlations}

Let $\boldsymbol{I}\equiv \epsilon \boldsymbol{I}_\epsilon\simeq \epsilon_k \boldsymbol{I}(\epsilon_k)/\Delta\epsilon_k$ be the \gama-ray brightness in LAT energy bin $k$ of mean energy $\epsilon_k$ and width $\Delta \epsilon_k$, with $\boldsymbol{I}(\epsilon_k)$ the energy intensity integrated over the bin.
Henceforth, sky maps over the $H$ HEALPix pixels $p$ in the region of interest (ROI) are denoted by bold symbols, and their pixel values by a subscript $p$.

A measure $I_\theta$ of the brightness correlated with an X-ray map $\boldsymbol{X}$ on angular scales smaller than $\theta$ is obtained by decomposing
\begin{equation}\label{eq:Template}
  \boldsymbol{I} = \langle \boldsymbol{I} \rangle + \boldsymbol{X}_\theta I_\theta + \boldsymbol{R}\coma
\end{equation}
where $\langle\ldots\rangle$ is an unweighted average over the ROI pixels and
\begin{equation}\label{eq:Xtheta}
  \boldsymbol{X}_\theta\equiv \left[\boldsymbol{X}-G(\boldsymbol{X})-a\right]/\langle \boldsymbol{X}\rangle
\end{equation}
is the small-scale X-ray contrast.
Here, $G$ is a Gaussian filter of radial dispersion $\theta=2\dgr$, chosen well above the angular scales of interest.
The constant $a$ is calibrated in each energy bin to maintain a zero weighted mean, $\llangle \boldsymbol{X}_\theta \rrangle=0$, where $\llangle\ldots\rrangle\equiv \langle \boldsymbol{\mathcal{E}}\ldots\rangle/\langle \boldsymbol{\mathcal{E}}\rangle$ is a sky average weighted by the LAT exposure $\boldsymbol{\mathcal{E}}$, the product of effective area and exposure time.

The residual $\boldsymbol{R}$ is defined to have zero exposure-weighted correlation with $\boldsymbol{X}_\theta$, \ie $\llangle \boldsymbol{X}_\theta\boldsymbol{R} \rrangle=0$.
Then
\begin{equation}\label{eq:Itheta}
  I_\theta = \frac{\mathrm{Cov}_{\mathcal{E}}(\boldsymbol{X}_\theta,\boldsymbol{I})}{\llangle \boldsymbol{X}_\theta^2 \rrangle} \coma
\end{equation}
where $\mathrm{Cov}_{\mathcal{E}}(\boldsymbol{Q}_1,\boldsymbol{Q}_2)\equiv\llangle \boldsymbol{Q}_1\boldsymbol{Q}_2\rrangle-\llangle \boldsymbol{Q}_1\rrangle\llangle \boldsymbol{Q}_2\rrangle$ is the exposure-weighted covariance, and the Poisson noise of the photon counts gives
\begin{equation}
  \mathrm{Var}(I_\theta) \simeq \frac{\epsilon_k/s_0}{H\delta\Omega \langle \boldsymbol{\mathcal{E}}\rangle} \frac{\llangle \boldsymbol{X}_\theta^2 \boldsymbol{I} \rrangle}{\llangle \boldsymbol{X}_\theta^2 \rrangle^2}\fin
\end{equation}
The physical results (spectral features, cross sections) would not change if one consistently replaces $\langle \boldsymbol{I}\rangle$ by $\llangle \boldsymbol{I}\rrangle$ (Eq.~\eqref{eq:Template} then giving $\llangle \boldsymbol{R}\rrangle=0$), or $\langle \boldsymbol{X}\rangle$ by $\llangle \boldsymbol{X}\rrangle$ (which would be unnatural for X-rays).

In the western Galactic hemisphere, high-resolution maps are available in several X-ray bands from the first eROSITA All-Sky Survey (eRASS1) \cite{MerloniEtAl24eRASS1}. We merge the survey tiles of each band into a full-sky HEALPix grid of order 10, for direct comparison with the \gama-ray data, and mask bright X-ray sources, which are not necessarily good tracers of LSS.
We choose the broad, $0.2$--$2.3\keV$ low-energy band, in which the underlying instrumental, non-X-ray background (NXB) is $\sim40\%$ of the diffuse signal \cite{ZhengEtAl24}.
The nominal map retains the NXB (Appendix \ref{app:eROSITA}); maps in this and other bands, with and without NXB removal, serve as robustness tests (\S\ref{subsec:Robustness}).
See Table \ref{tab:CorrParameters} for a parameter summary, and Appendix \ref{app:eROSITA} for additional details on the X-ray data reduction.

To isolate narrow $\delta U \equiv U-\mathrm{GPR}(U)$ spectral features of a given spectrum $U$, we use Gaussian process regression (GPR) on energy scales greatly exceeding the instrumental energy resolution, to model and remove the smooth astronomical continuum along with the broad leakage of the PSF and energy dispersion.
The non-parametric, widely used\citep{AlcantaraEtAl15, KaramanavisEtAl16, AgarwalEtAl25} GPR \cite{RasmussenWilliams06} results are robust (\S\ref{subsec:Robustness}); for details, including a final linear detrend, see Appendix \ref{app:GPR}.

We adopt the detrended X-ray-correlated spectrum $\delta I_\theta$ as our primary spectrum, as it isolates the component traced by the X-rays on angular scales smaller than $\theta$.
The detrended ambient spectrum $\delta I_0\equiv \delta\llangle \boldsymbol{I}\rrangle$ serves as a control.
The LSS-enhanced sum $\delta I\equiv\delta (\llangle \boldsymbol{I}\rrangle+I_\theta)$, \ie the modeled brightness of a pixel of unit contrast $\boldsymbol{X}_\theta=1$,
adds to $\delta I_\theta$ the ambient spectrum, which shows no correlated spectral features (\S\ref{subsec:Robustness}) and mainly adds noise.
The analysis of $\delta I$ yields similar results, presented in Appendix \ref{app:dI}.

\subsection{Matched spectral filters}
\label{subsec:MatchedFilters}

An inclusive (averaged over event types), effective-area-weighted \cite{AckermannEtAl13, LiangEtAl16, Gammapy2023} IRF is synthesized from the official \texttt{P8R3\_ULTRACLEANVETO\_V2} calibration\citep{BruelEtAl18Pass8R3} matrices\footnote{\url{https://heasarc.gsfc.nasa.gov/FTP/fermi/calib_data/lat/}}, applicable also to \texttt{P8R3\_ULTRACLEANVETO\_V3} (Appendix \ref{app:IRF}).
The probability density $P(s,\epsilon)$ of the fractional energy error $s\equiv(\epsilon'-\epsilon)/\epsilon$, where $\epsilon'$ is the reconstructed energy, is tabulated (as a 15-parameter fit formula) for each of the four energy-dispersion (EDISP) event-type quartiles $q$ (ranking events by the quality of their energy reconstruction)\cite{AjelloEtAl16Spectral}, on a grid of eight incidence angles $\vartheta$ by 23 energies $\epsilon$ of the incident photon. This density is averaged over $q$ and over $\cos\vartheta$, and interpolated over $\ln\epsilon$.
For our wide-sky analyses, this averaging is weighted by the effective area, itself tabulated for each $q$ on a finer 32-by-74 grid of $\cos\vartheta$ and $\ln\epsilon$.
See Appendix \ref{app:IRF} for details.

\subsubsection{Line filters}

To search for narrow spectral features, the interpolated IRF is applied as a sliding aperture (matched filter) across the logarithmically binned, detrended, differential brightness spectrum $\delta I_\theta(\epsilon_k) = y_k\pm\sigma_k$, at bin energies $x_k=\ln\epsilon_k$.
For a putative signal at energy $\epsilon_0$, the total line brightness $\mathcal{I}$ (henceforth, $\mathcal{I}$ denotes a brightness integrated over a line or a band) and its local $Z$-score $Z_l$ are evaluated using a least-squares estimator,
\begin{equation} \label{eq:MatchedFilter}
\mathcal{I}(\epsilon_0) = \frac{\sum_k w_k y_k}{\sum_k w_k^2} \, , \quad Z_l(\epsilon_0) = \frac{\sum_k w_k y_k}{\sqrt{\sum_k w_k^2 \sigma_k^2 }} \, ,
\end{equation}
where the interpolated $w_k \equiv P(\epsilon_k/\epsilon_0 - 1, \epsilon_0)$, the averaged IRF (Appendix~\ref{app:IRF}), is normalized (to unit integral over the logarithmic $x$) over the full grid and truncated outside the nominal $[0.25, 1.50]\epsilon_0$ range to suppress extrapolation artifacts.

Local $Z_l$ maxima are identified
on a fine logarithmic grid of $0.2\%$ steps in $\epsilon_0$.
Interval limits $\epsilon_0\pm\Delta\epsilon_\pm$ around each maximum are inferred on this grid as asymmetric profile-likelihood bounds, where the test statistic $\textrm{TS}\equiv Z_l^2(\epsilon_0)$ falls by $\Delta\textrm{TS}=1$ below its local maximum.
These intervals reflect the likelihood geometry and are not standard Gaussian uncertainties, but mock tests show that they contain the injected energy in $\sim70\%$
of cases (see Appendix \ref{app:Mock}).

Multiple lines are extracted iteratively, in decreasing order of $Z_l$: a maximum at $\epsilon_0$ is extracted only if its window $[1-f_L, 1+f_H]\epsilon_0$ does not overlap a previously excluded window, and this window is then excluded, to prevent overlapping or duplicate detections.
A symmetric window with nominal $f_L=f_H=f=0.1$ roughly accounts for the instrumental energy-dispersion tails in the $10$--$100\GeV$ energy range of interest; a smaller $f=0.03$ is also examined.

\subsubsection{Line complexes}

Energy-momentum conservation dictates that two WIMPs (not necessarily of the same species) of mean mass-energy $E_{\chi}\equiv \overline{m}_\chi c^2$, annihilating into a neutral standard-model boson $\varphi$ ($\gamma$, $Z$, or $h$) and a photon, produce an emission line at
\begin{equation} \label{eq:kin}
\epsilon \simeq E_{\chi}-\frac{m_\varphi^2 c^4}{4E_{\chi}} \coma
\end{equation}
so the triad is fixed by a single parameter, $\overline{m}_\chi$.
The $\gamma\gamma$ channel thus includes two $\epsilon\simeq E_\chi$ photons, accompanied by a lower-energy $\gamma Z$ line (if $E_\chi>m_Z c^2/2\simeq 45.6\GeV$) and an even lower-energy $\gamma h$ line (if $E_\chi>m_h c^2/2\simeq 62.6\GeV$).
We adopt the central values of present Z-boson $m_Zc^2=91.1880\pm0.0020\GeV$ and Higgs $m_hc^2=125.20\pm0.11\GeV$ mass estimates\cite{PDG2024},
and convolve the $\gamma Z$ channel with the $\sim2.5\GeV$ (width of $m_Zc^2$ in Eq.~\eqref{eq:kin}) Breit--Wigner resonance.

When scanning for a kinematically constrained $n$-line complex, such as an $n=3$ triad of $\gamma\gamma$, $\gamma Z$, and $\gamma h$ WIMP annihilation channels, the composite local significance becomes the unweighted sum
\begin{equation}\label{eq:CoAddedZ}
  Z_{l} = n^{-1/2}\sum_{i=1}^n Z_{l,i}\coma
\end{equation}
aggregated across the $n$ lines convolved with the IRF, assuming independent background fluctuations across the channels.
(Any residual correlations between channels, \eg induced by the GPR detrending, are captured by the shuffle calibration; see Appendix \ref{app:MC}.)
The triad channel $Z_{l,i}$ scores are denoted $Z_{\gamma\gamma}$, $Z_{\gamma Z}$, and $Z_{\gamma h}$.
A line complex is scanned over its free masses (one for a triad, and two for the nonads defined below),
which equal the energies of its independent $\gamma\gamma$ lines; the other channels follow from Eq.~\eqref{eq:kin} kinematics in a rigid template, so exclusion windows (with a narrow $f=0.015$ for triads) are applied only around the independent lines.

Such a triad filter is sensitive to $\overline{m}_\chi c^2$ in the range $\sim64$--$100\GeV$ (where all three lines lie within our $3$--$100\GeV$ search window), and is scanned in $0.1\%$ steps in $\overline{m}_\chi$.
Nonads, formed by the two self-annihilation triads of WIMPs of masses $m_1$ and $m_2$ and their intermediate, cross-annihilation triad of mean mass $(m_1+m_2)/2$, are similarly scanned, in $0.5\%$ steps in each mass.
A nonad scan uses a two-parameter composite aperture, which sums the nine line apertures with equal integrated brightness.

Nonads are scanned over pairs whose triads are separated by more than a window $f_t=0.02$, \ie $m_2/m_1>(1+f_t)/(1-f_t)\simeq1.04$, with $f=0.05$ separating independent nonads.
The $f_t$ window separates the two self-annihilation triads from each other, but not from the intermediate, cross-annihilation triad.
The same triad may thus serve as a self-annihilation triad of one nonad and as the cross-annihilation triad of another, allowing alternative nonad readings of the same triads.

Summing the nine apertures, which essentially fits a common brightness to all lines, is the simplest way to keep the composite $Z$-score correctly normalized for the overlapping nonad lines, whose $Z_{l,i}$ are strongly correlated.
The plain sum of these $Z_{l,i}$, Eq.~\eqref{eq:CoAddedZ} with $n=9$, would overestimate $Z_l$ by a factor $\sim1.2$--$1.8$.
An alternative equal-$Z$ weighting, as for triads but with the normalization corrected for the overlapping lines, gives very similar nonad $Z_l$
(Pearson correlation coefficient $0.99$ over all
scanned mass pairs; $Z_l$ typically lower by $\lesssim0.1$ for $Z_l>2.5$) and global $Z$-scores, with only minor variations, such as the ranking of very similar maxima.

Triad and nonad masses carry $\Delta\textrm{TS}=1$ bounds analogous to those of lines.
For a nonad, the bound on each mass is profiled, maximizing $Z_l$ over the other mass at each trial value, and interpolated linearly between the $0.5\%$ grid points.

\subsubsection{Redshift broadening}

To test for emission from a range of redshifts, consider line emission spread over redshifts $z$, measured relative to the rest frame in which the aperture is centered, with a line-of-sight emissivity $\smash[t]{\tilde{\zeta}}(z)$
per unit $z$ over a range $[\Delta z_{-},\Delta z_{+}]$.
Then $P(s,\epsilon)$ generalizes to
\begin{equation}\label{eq:Dz}
\int_{\Delta z_{-}}^{\Delta z_{+}} (1+z) P\left[ (1+s)(1+z) - 1, \frac{\epsilon}{1+z} \right] \tilde{\zeta}(z) \, dz \coma
\end{equation}
normalized to unity when integrated over $s$.
For simplicity, define the redshift interval $\{\Delta z_-,\Delta z_+\}=\{\min(\Delta z,0),\max(0,\Delta z)\}$ as a function of a single parameter $\Delta z$.

A one-dimensional $\Delta z$ scan then serves to approximately probe both the redshift distribution of the signal (\S\ref{subsec:Dz}) and the spectral structure adjacent to the channels of a line complex (just below them for $\Delta z>0$ and just above them for $\Delta z<0$).
We nominally adopt $\smash[t]{\tilde{\zeta}}(z)=\const.$, which is convenient for such a scan but less relevant for true redshift broadening, which should rapidly drop with $|z|$; for example, a $\smash[t]{\tilde{\zeta}}$ peaking at $z\simeq0.03$ and declining as $z^{-2}$ beyond it is more natural for the correlation signal.
Appendix \ref{app:PeakZ} models the redshift distribution $\mathcal{Z}(z)$ of the signal anticipated from LSS, for $p$-wave annihilation in Navarro--Frenk--White\cite{NFW97} (NFW) halos of a standard mass function.

\subsubsection{Global significance}
\label{subsubsec:GlobalSignificance}

A matched-filter scan tests many energies or masses, so noise alone yields some elevated local $Z_l$ values.
Global $Z$-scores, which correct for these multiple trials, are estimated by a few different methods.

Our nominal calibration is based on random shuffling of the X-ray contrast map $\boldsymbol{X}_\theta$, producing a null distribution of (nominally)
$2\times 10^4$ realizations.
Each such shuffled-contrast sky permutes the ROI pixels of the X-ray contrast $\boldsymbol{X}_\theta$
(with the constant $a$ recalibrated), with no change to the LAT photon counts and exposure of each pixel.
The spectrum is then rebuilt, detrended, and scanned in the same way as the real data.

Each shuffled-contrast sky thus retains any non-Gaussian or spatially correlated noise in the \gama-ray data, but not their correlation with the X-ray map. The \gama-ray emission correlated with the real contrast
projects onto the random shuffled contrast with zero mean
and a relative standard deviation $\sim H^{-1/2}\simeq7\times10^{-4}$.

The $I_\theta$ spectrum of a shuffled-contrast sky thus lacks the smooth continuum of the data.
As the GPR kernel amplitude is fitted to each spectrum (Appendix~\ref{app:GPR}), a spectrum lacking this continuum would be detrended differently from the data.
We therefore add to $I_\theta$ of each shuffled-contrast sky, before detrending, the continuum that detrending fitted to the real data.
This continuum is smooth on the GPR scale and carries no bin-to-bin noise, so each sky retains its own photon counts and errors, and mimics the data in the absence of narrow line emission.
The sum $\llangle \boldsymbol{I}\rrangle+I_\theta$ (\ie $\delta I$ before detrending) of each sky then also carries the full continuum of the data, as the ambient spectrum is unaffected by shuffling.

Spatially correlated \gama-ray fluctuations add slightly less noise to the spectra of shuffled-contrast skies than to the data, as the shuffled contrast lacks the spatial coherence of the real contrast. However, the standard deviation of the normalized $\delta I_\theta$ residuals is only $\sim1.03$ (see \S\ref{sec:Results}), limiting this difference to a few percent.
The $Z_l$ of each shuffled-contrast sky is normalized by its own propagated errors, as for the data.
Normalizing the skies instead by the slightly larger errors of the data would narrow the null distribution
and raise the calibrated significance by $\sim0.1$--$0.3\sigma$, so the adopted normalization is the more conservative choice.

For each shuffled-contrast sky and each type of scan (lines, triads, or nonads), we record the highest $Z_l$ among the local maxima of the scan.
Local maxima at an edge of the scanned range, where $Z_l$ still rises towards the edge suggesting that the actual peak lies outside the range, are excluded, both in the shuffled-contrast skies and in the data.
The global $p$-value $p_g$ of a feature with local $Z_l$ score is defined as the fraction of shuffled-contrast skies whose recorded value, for the same type of scan, exceeds $Z_l$, and $Z_g$ is defined as the one-sided Gaussian equivalent of $p_g$.
Weaker features of a scan, such as the second triad, are calibrated against the same distribution as the strongest one.

As a consistency check,
we also simulate $10^5$ spectra of independent Gaussian noise, with zero mean and, in each energy bin $k$, a standard deviation equal to the propagated error $\sigma_k$ of the tested spectrum ($\delta I_\theta$ or $\delta I$).
These spectra, which have no continuum, are scanned in the same way as the data, but without detrending.
Monte Carlo $p$-values are computed as for the shuffles, and their one-sided Gaussian equivalent is the Gaussian-noise global $\smash{Z_g^{(G)}}$ score.

Every $p$-value, $Z$-score, or fraction quoted from shuffled-contrast skies, Gaussian noise spectra, or mocks rests on at least $10$ (typically $>30$) realizations meeting the tested condition.

For reference, we also quote the analytic \smash{\v{S}id\'{a}k}\cite{Sidak1967} estimate $\smash{Z_g^{(a)}}$ based on $1 - p_{g} \simeq (1 - p_{l})^{N_t}$, where $p_l$ is the local $p$-value and $N_t$ is the effective number of independent trials, taken as the logarithmic search range divided by the logarithmic exclusion width $f_L+f_H$, within which features are not independent; for the two-mass nonad scan, $N_t\to N_t^2/2$, as $m_1<m_2$.
For triads, an independent analytic estimate $\smash{Z_g^{(n)}}$ is obtained from the $p_{g} \simeq p_{l} + N(Z_0)\exp[-(Z_l^2 - Z_0^2)/2]$ Euler characteristic up-crossing approximation \cite{GrossVitells10}, where the mean number $N(Z_0)$ of background fluctuations exceeding a $Z_0=1$ threshold is computed from the IRF widths of the three channels as they shift with $\overline{m}_\chi$.

The four estimates, $Z_g$, $\smash{Z_g^{(G)}}$, $\smash{Z_g^{(a)}}$, and $\smash{Z_g^{(n)}}$, are compared, and their differences traced, in Appendix~\ref{app:MC}.
Significance levels written as $N\sigma$ are Gaussian-equivalent and one-sided; for blind scans they are global, and for the externally fixed masses of \S\ref{sec:43GeV}, $Z_g$ and $\smash{Z_g^{(G)}}$ denote the calibrated significance of the single test.
Appendix~\ref{app:MC} also specifies the null distributions of the fixed-mass tests.

\subsection{Inferring $\overline{\sigma v}_p$ from X-ray correlations}
\label{subsec:SigmaCorrelations}

The ratio between \gama-ray emissivity $j_\gamma$ of WIMP annihilation into channel $i$ photons, and X-ray emissivity $j_X$ of thermal bremsstrahlung, may be written as
\begin{equation}
    \frac{j_\gamma}{j_X} = \frac{Y_i \epsilon_i \beta^2\overline{\sigma v}_{p,i}/2\eta_\chi}
    {\Lambda(z)/(\mu_e \mu_H)} \left( \frac{m_p \Omega_\chi b_\chi}{m_\chi \Omega_b} \right)^2 \, ,
\end{equation}
where $\mu_e\simeq \mu_{e,0}\equiv 1.17$ and $\mu_H\simeq \mu_{H,0}\equiv 1.40$ are the mean molecular weights per free electron and per hydrogen atom, respectively, $m_p$ is the proton mass, $\Omega_b\simeq0.049$, $\Omega_c\simeq0.265$, and $\Omega_\chi\le\Omega_c$ are the baryon, cold DM, and WIMP density parameters\cite{AghanimEtAl20Planck}, $\Lambda(z)$ is the rest-frame band-limited X-ray cooling function, $\epsilon_i$ is the energy of channel-$i$ photons, and $\eta_\chi=1$ ($\eta_\chi=2$) for Majorana (Dirac) WIMPs.
The local DM boost factor $b_\chi(\boldsymbol{r}) \equiv \rho_\chi(\boldsymbol{r})\Omega_b/[\rho_b(\boldsymbol{r})\Omega_\chi]$, where $\rho_b$ is the baryon mass density, is uniformly 1 for a perfect overlap of baryons and DM.

Integrating the \gama-ray and X-ray emissivities along the line of sight $\ell$ yields their brightness ratio
\begin{equation}\label{eq:RatioA}
    \mathcal{R}_i \simeq  \frac{Y_i \epsilon_i \overline{\sigma v}_{p,i}}{2\eta_\chi{\mathscr L}} \left( \frac{m_p \Omega_\chi}{m_\chi \Omega_b} \right)^2 \coma
\end{equation}
where the effective cooling function
\begin{equation}\label{eq:Llos}
    {\mathscr L} = \frac{\int (1+z)^{-4}(\mu_e \mu_H)^{-1} \Lambda(z) \rho_b^2\, d\ell}{\int (1+z)^{-4} \beta^2 b_\chi^2 \rho_b^2\, d\ell}
\end{equation}
encapsulates the integrated structure parameters along with the respective cosmological dimming factors.

In order to tie the underlying cross sections to the observed cross-correlation, we replace ${\mathscr L}$ by its sky average with $(\boldsymbol{X}_\theta\boldsymbol{\mathcal{E}})$ weights,
\begin{equation}\label{eq:L}
    {\mathscr L}\equiv \frac{\llangle \boldsymbol{X}_\theta\int (1+z)^{-4}(\mu_e \mu_H)^{-1} \Lambda(z) \rho_b^2\, d\ell \rrangle}{\llangle \boldsymbol{X}_\theta\int (1+z)^{-4} \beta^2 b_\chi^2 \rho_b^2\, d\ell \rrangle} \fin
\end{equation}
Indeed, $\mathcal{R}_i$ can now be related to the ratio between channel brightness $\mathcal{I}_i=\mathbb{M}_i(\delta I_\theta)$ and mean X-ray brightness $\langle \boldsymbol{X}\rangle$, which converts the contrast $\boldsymbol{X}_\theta$ to X-ray brightness,
\begin{equation}\label{eq:RatioNL}
    \mathcal{R}_i \simeq \frac{\mathcal{I}_i}{\langle \boldsymbol{X} \rangle}
    \simeq \frac{\mathbb{M}_i\left[\delta\,\mathrm{Cov}_{\mathcal{E}}(\boldsymbol{X}_\theta,\boldsymbol{I})\right]}{\llangle\boldsymbol{X}_\theta \boldsymbol{X}\rrangle} \coma
\end{equation}
where $\mathbb{M}_i$ is the matched filter of Eq.~\eqref{eq:MatchedFilter} evaluated at $\epsilon_0=\epsilon_i$,
and the second equality neglects the
$\sim0.6\%$ correlation between $\boldsymbol{X}_\theta$ and $G(\boldsymbol{X})$.
This equation does not account for losses, discussed in \S\ref{subsec:Losses}.

It is useful to define the dimensionless factor
\begin{equation}\label{eq:W}
  D\equiv \left(\frac{\Omega_c}{\Omega_\chi}\right)^2 \frac{\mu_{e,0}\mu_{H,0}\eta_\chi}{10^{5}\Lambda(0)} {\mathscr L} \coma
\end{equation}
which collects the unknown dark-sector and structural parameters.
We normalize $D$ to unity for the $z=0$ annihilation signal from DM composed of a single Majorana species that traces the nominal low-redshift intracluster medium (ICM) of $\beta^2=10^{-5}$ clusters.
Equating Eqs.~\eqref{eq:RatioA} and \eqref{eq:RatioNL} facilitates a measurement of the intrinsic $p$-wave cross section normalized by this ignorance factor,
\begin{equation}\label{eq:CrossSection}
    \frac{\overline{\sigma v}_{p,i}}{D} \simeq \left( \frac{m_\chi \Omega_b}{m_p \Omega_c} \right)^2 \frac{2\times10^{5}\Lambda(0)}{\mu_{e,0}\mu_{H,0}\,\epsilon_i Y_i}\,\mathcal{R}_i \fin
\end{equation}
Finally, the DM boost factor may be written as
\begin{equation}\label{eq:B}
  B\equiv \frac{D(b_\chi=1)}{D} = \frac{\llangle \boldsymbol{X}_\theta\int (1+z)^{-4}\beta^2 b_\chi^2 \rho_b^2 \, d\ell \rrangle}{\llangle \boldsymbol{X}_\theta\int (1+z)^{-4}\beta^2 \rho_b^2 \, d\ell \rrangle}\coma
\end{equation}
normalized to the clumpy $\beta^2 \rho_b^2$, unlike the conventional normalization to a smooth $\beta^2\rho_\chi^2$ model.

The band-limited cooling function is estimated\cite{SutherlandDipotal93, SmithEtAl01} as $\Lambda(z)\simeq\Lambda(0)\simeq 8.5\times 10^{-24}\erg\cm^3\se^{-1}$ for a typical ICM of temperature $4$ keV and metallicity $0.3 Z_\odot$; $\Lambda$ is not sensitive to the exact choice of ICM parameters.
The mean X-ray brightness of the nominal map (including the NXB), averaged over the unmasked pixels, is $\langle \boldsymbol{X}\rangle\simeq 9.0\times 10^{-8}\erg\se^{-1}\cm^{-2}\sr^{-1}$.
Relevant values are summarized in Table \ref{tab:CorrParameters}.

The intrinsic cross section $\overline{\sigma v}_p$, defined in Eq.~\eqref{eq:EffCS} and
used for the previous (continuum) \gama-ray upper limits quoted in \S\ref{sec:Intro}, refers here, without an index, to the primary, $\chi\chi \to \gamma\gamma$ line channel, of $Y=2$ photon multiplicity.
Such estimates convert to the cross section $\overline{\sigma v}_{p,i}$ of another channel $i$, for the same number of line photons, by a factor $2/Y_i$.
Assuming an isotropic halo, the single-particle 3D velocity dispersion is $3\beta^2$, and the 3D relative velocity dispersion becomes $6\beta^2$.
While keeping the definition \eqref{eq:EffCS} for comparison with the literature, note that the implied $\overline{\sigma v}_p$ is six times larger than for a more natural definition based on the 3D relative velocity.

\subsection{Biases and losses}
\label{subsec:Losses}

Photon energies are quoted as observed, and WIMP masses as inferred from them, so both are biased low by the emission redshift.
To quantify this bias for the lines found here, mock single lines emitted at the energies of $L_{43}$ and of the $\gamma\gamma$ lines of the two leading triads ($43.2$, $67.5$, and $71.9\GeV$) are spread over the LSS redshift distribution $\mathcal{Z}(z)$ of Appendix~\ref{app:PeakZ} and injected into shuffled-contrast skies as in Appendix~\ref{app:Mock}.
The line scan recovers them $\sim5$--$6\%$ (median) below their emitted energy, consistent with the median $z\simeq 0.06$--$0.07$ of $\mathcal{Z}(z)$.

Triad and nonad mass offsets are smaller because the $\gamma Z$ and $\gamma h$ energies vary logarithmically faster than their $\gamma\gamma$ counterpart (by factors $\sim2.3$ and $\sim7$ for $\overline{m}_\chi c^2\simeq72\GeV$), thus pinning the mass.
However, the overall effect depends on the relative $Z$-scores of the lines.
Mock triads, spread over $\mathcal{Z}(z)$ and injected as above at the masses of the two leading triads, each with its channel brightness ratios taken as measured, are recovered by the triad scan with a median only $\sim0.9\%$ below their mass.
The measured ratios carry $\sim80\%$ statistical uncertainties and are altered by the pipeline, thus modifying the offset by a factor $\lesssim 2$; the offset would approach the single-line shift for triads whose $Z_l$ is dominated by their $\gamma\gamma$ line.
The offset is revisited in \S\ref{sec:43GeV}.

Only a fraction $\mathcal{F}_{\mbox{\tiny PSF},i}$ of the X-ray-correlated \gama-ray signal in line $i$ survives the spatial smoothing by the LAT PSF, which spreads line photons away from the X-ray-bright pixels.
Of the remainder, the pipeline, including the matched filter, measures on average only a fraction $\mathcal{F}_i$, which depends on the signal strength (see below).
In X-rays, only a fraction $\mathcal{F}_b$ of the contrast variance $\llangle \boldsymbol{X}_\theta^2\rrangle$ is due to LSS bremsstrahlung.
Accounting for these loss fractions, Eq.~\eqref{eq:RatioNL} becomes
\begin{equation}\label{eq:Ratio}
    \mathcal{R}_i \simeq \frac{\mathcal{I}_i/(\mathcal{F}_i\mathcal{F}_{\mbox{\tiny PSF},i})}{\mathcal{F}_b\langle \boldsymbol{X} \rangle}
    \simeq \frac{\mathbb{M}_i\left[\delta\,\mathrm{Cov}_{\mathcal{E}}(\boldsymbol{X}_\theta,\boldsymbol{I})\right]}{\mathcal{F}_i\mathcal{F}_{\mbox{\tiny PSF},i}\mathcal{F}_b\llangle\boldsymbol{X}_\theta \boldsymbol{X}\rrangle} \coma
\end{equation}
so Eq.~\eqref{eq:CrossSection} gives the loss-corrected cross section
\begin{equation}\label{eq:CrossSectionLosses}
    \frac{\overline{\sigma v}_{p,i}}{D} \simeq  \left( \frac{m_\chi \Omega_b}{m_p \Omega_c} \right)^2 \frac{2\times10^{5}\Lambda(0)}{\mu_{e,0}\mu_{H,0}\,\epsilon_i Y_i}\, \frac{\mathcal{I}_i}{\mathcal{F}_i\mathcal{F}_{\mbox{\tiny PSF},i}\mathcal{F}_b\langle \boldsymbol{X} \rangle} \fin
\end{equation}

Contributions to $\llangle \boldsymbol{X}_\theta^2\rrangle$ that do not directly trace the gas, such as X-ray Poisson noise, unidentified AGN, or masking residuals, render $\mathcal{F}_b<1$.
As $\mathcal{F}_b$ relies on the halo model, and is bounded only in combination with $\mathcal{F}_{\mbox{\tiny PSF},i}$ (see below), loss-corrected cross sections are quoted for $\mathcal{F}_b=1$, and are thus underestimated by a factor $1/\mathcal{F}_b$.

Cross sections are quoted and shown in figures as measured, \ie inferred using Eq.~\eqref{eq:CrossSection} from the measured line brightness, unless stated as loss-corrected and thus inferred from Eq.~\eqref{eq:CrossSectionLosses}.
The loss factors are estimated by different means: $\mathcal{F}_i$ using mocks (below), $\mathcal{F}_{\mbox{\tiny PSF},i}$ with the halo model of Appendix~\ref{app:PeakZ}, and $\mathcal{F}_b$ by smoothing the X-ray map with the LAT PSF, which gives an upper bound on $\mathcal{F}_b\mathcal{F}_{\mbox{\tiny PSF},i}$ (with a small correction from the halo model) and, combined with the halo model, an estimate of $\mathcal{F}_b$.
Appendix~\ref{app:LossFactors} details such estimates, listing the various loss factors of the nominal triad for different assumptions on the redshift distribution.

The pipeline loses line brightness mainly in two ways: its $z=0$ apertures miss part of the redshift-spread signal, and the detrending absorbs part of each line.
The fraction $\mathcal{F}_i$ captures these two losses, and the mean effect of the noise, albeit not all other losses and systematics.
For instance, template mismatches due to errors in the LAT energy scale or IRF are not captured.

We estimate $\mathcal{F}_i$ with mocks of the nominal triad, processed by the same pipeline as the data.
These mocks, spread in redshift over $\mathcal{Z}(z)$ or variations thereof, are added to the correlated spectrum $I_\theta$ of each of $200$ shuffled-contrast skies.
Shuffling keeps the actual \gama-ray background and its noise, but removes its correlation with the contrast, so $I_\theta$ of a shuffled-contrast sky lacks any real signal.
As in the case of calibration (\S\ref{subsubsec:GlobalSignificance}), each sky receives, before detrending, the continuum of the data, which raises the fraction of a line absorbed by detrending (Appendix~\ref{app:Mock}).

Each mock-based spectrum is then detrended with errors estimated from its own photon counts, as are the real data, and the brightness of each channel is measured with the matched filter $\mathbb{M}_i$ of the nominal triad.
These mocks, added as spectra without PSF smoothing, lack PSF losses, which are accounted for separately by $\mathcal{F}_{\mbox{\tiny PSF},i}$.
They also lack the photons of the signal itself, but this effect is small as the count-based data errors are not significantly elevated at the channel energies (Appendix~\ref{app:LossFactors}).

The fraction $\mathcal{F}_i$ depends on the signal strength, as the signal non-linearly affects the detrending, and as the pipeline measures a positive mean brightness even without a signal (the null mean; Appendix~\ref{app:MC}).
Triads are therefore injected at several strengths, and each $\mathcal{F}_i$ is evaluated at the strength that reproduces, on average, the measured brightness $\mathcal{I}_i$.
For the nominal triad, we find $\mathcal{F}_i\simeq\{0.30, 0.35, 0.34\}$ for the $\{\gamma\gamma,\gamma Z,\gamma h\}$ lines spread over $\mathcal{Z}(z)$, and $\mathcal{F}_{0,i}\simeq\{0.77, 0.75, 0.82\}$ for lines emitted at $z=0$, losing brightness only to detrending.
Appendix~\ref{app:LossFactors} lists $\mathcal{F}_i$ also for other redshift distributions, and discusses its dependence on mass.
Nonads are not loss-corrected, as their aperture fits one common brightness to all nine lines (which need not be equally bright).

The PSF reduces the contribution of each halo to the correlation signal through the limited overlap of its X-ray profile with its PSF-smoothed \gama-ray profile.
Averaged over $\mathcal{Z}(z)$ halos, the LAT PSF retains a modeled fraction $\mathcal{F}_{\mbox{\tiny PSF},i}\simeq\{0.52, 0.51, 0.46\}$ of the X-ray-correlated signal relative to a point-like PSF.
Larger fractions are recovered for redshift distributions that favor nearby, better-resolved halos (Appendix~\ref{app:LossFactors}).

We probe $\mathcal{F}_b$ by smoothing the X-ray map itself with the LAT PSF (at the energy of line $i$), and measuring the fraction $\mathcal{F}_{X,i}$ of its covariance with the contrast that survives the smoothing.
Smoothing should retain a fraction $\mathcal{F}_{\mbox{\tiny gas},i}$ of the gas contribution to this covariance, which we approximate with the halo model,
and a fraction $\mathcal{F}_{\mbox{\tiny noise},i}$ of the pixel-scale noise contribution,
which we measure by shuffling the contrast; hence, $\mathcal{F}_{X,i}\simeq\mathcal{F}_b\mathcal{F}_{\mbox{\tiny gas},i}+(1-\mathcal{F}_b)\mathcal{F}_{\mbox{\tiny noise},i}$.
Using $\mathcal{F}_{X,i}\simeq\{0.30, 0.28, 0.22\}$, $\mathcal{F}_{\mbox{\tiny gas},i}\simeq\{0.48, 0.47, 0.42\}$, and $\mathcal{F}_{\mbox{\tiny noise},i}\simeq\{0.25, 0.23, 0.16\}$ for the $\{\gamma\gamma,\gamma Z,\gamma h\}$ lines of the nominal triad, gives $\mathcal{F}_b\simeq\{0.22, 0.22, 0.21\}$, nicely consistent across channels.
These estimates depend on the absolute $\mathcal{F}_{\mbox{\tiny gas},i}$ scale of the halo model, so the limit $\mathcal{F}_b\mathcal{F}_{\mbox{\tiny PSF},i}\lesssim(\mathcal{F}_{\mbox{\tiny PSF},i}/\mathcal{F}_{\mbox{\tiny gas},i})\mathcal{F}_{X,i}\simeq1.1\,\mathcal{F}_{X,i}\simeq \{0.33,0.31,0.24\}$ is more robust.

\section{Results}
\label{sec:Results}

\begin{figure}[h!]
    \includegraphics[width=\columnwidth]{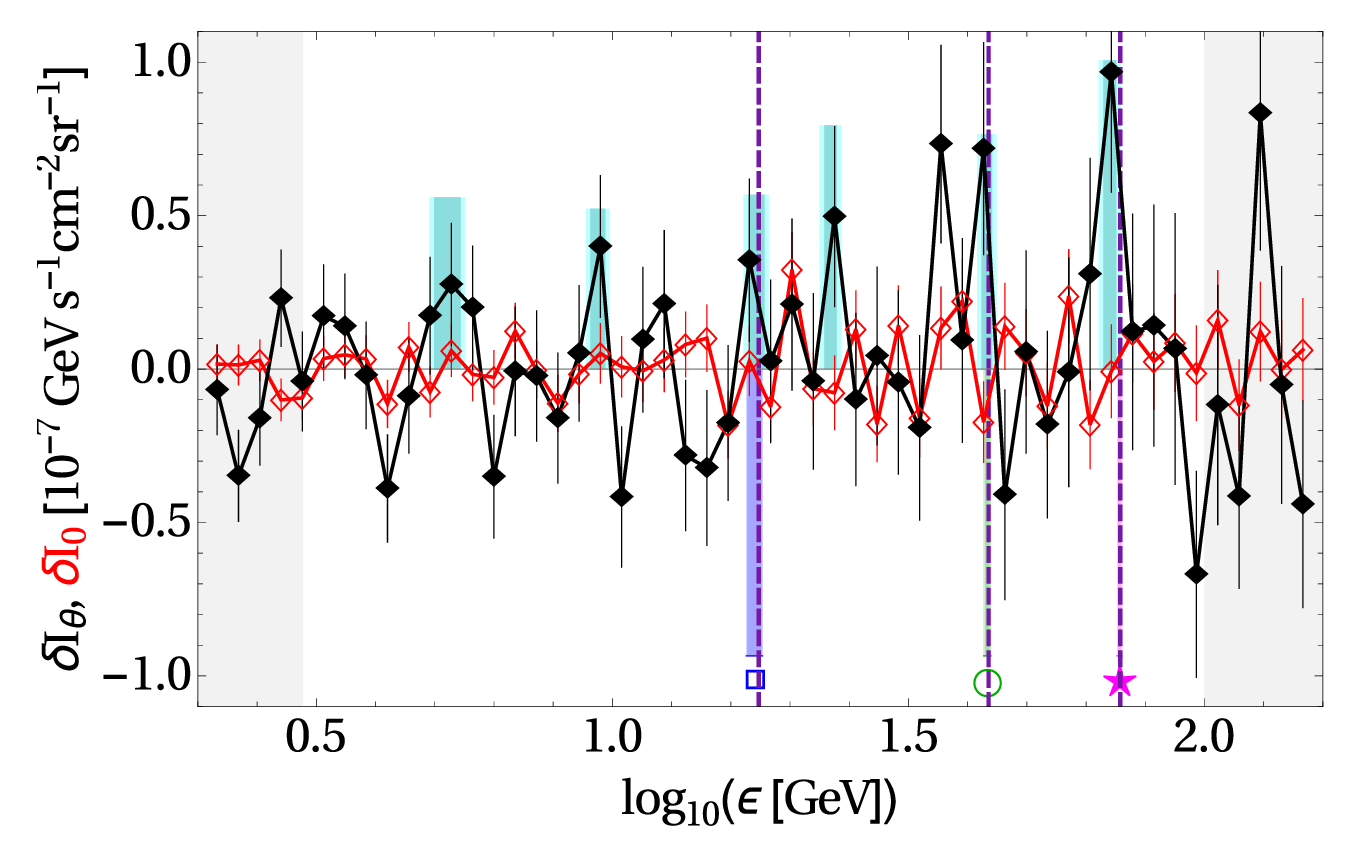}\\
    \includegraphics[width=\columnwidth]{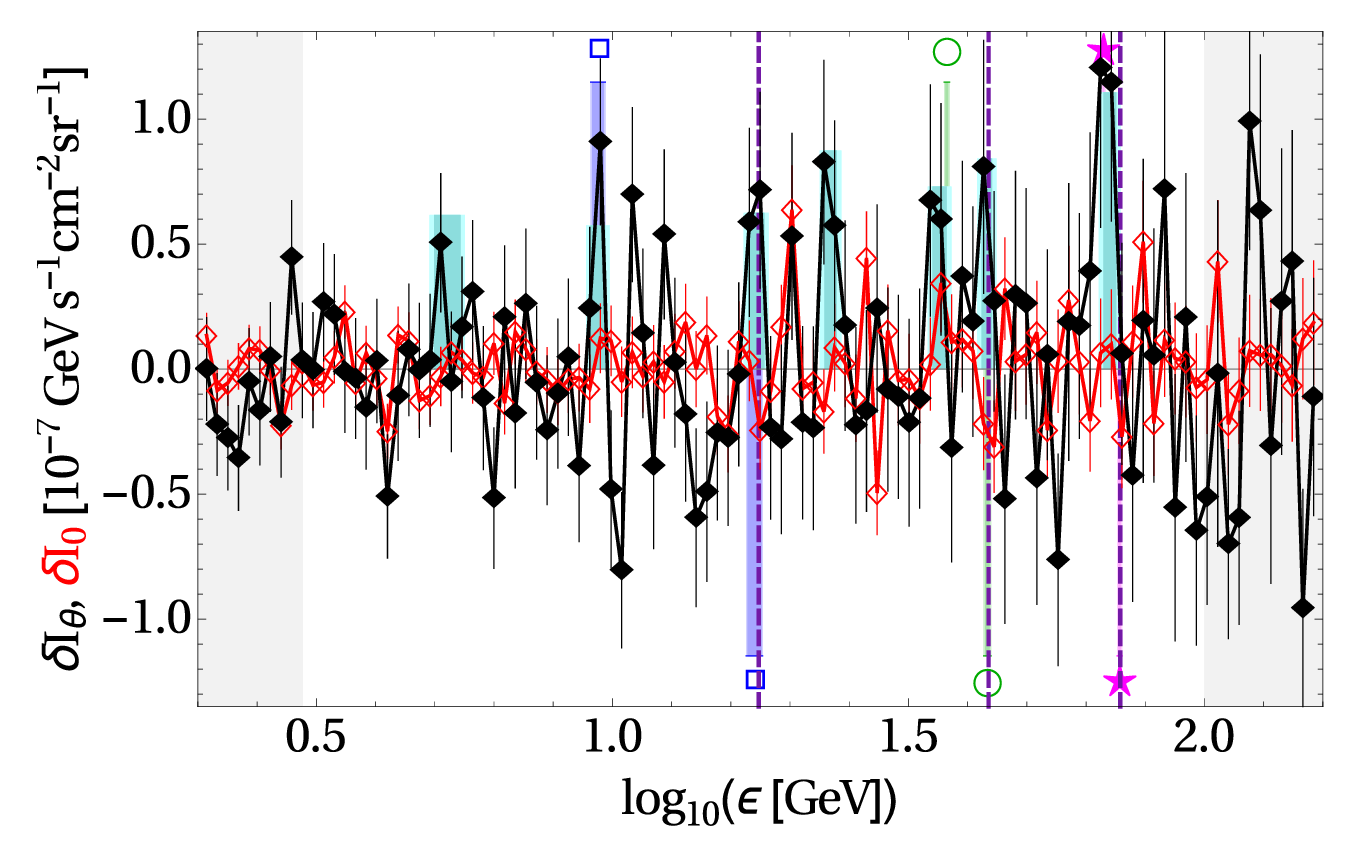}
	\caption{\label{fig:Spectra}
        X-ray-correlated LAT spectrum $\delta I_\theta$ (filled black diamonds), compared to the ambient spectrum $\delta I_0$ (empty red diamonds), rebinned at low ($s\simeq4s_0\simeq8.4\%$;
        $f=0.1$ line exclusion; top panel) or medium ($s\simeq2s_0\simeq4.2\%$; $f=0.03$; bottom panel) energy resolutions.
        Also shown are matched-filter lines with $Z_l>1$ (independent of rebinning; cyan bars of height $\propto Z_l$ and width $\Delta\epsilon_{\pm}$), and the three rest-frame channel energies of the nominal triad (\S\ref{sec:43GeV}; dashed purple).
        Annihilation channels $\gamma\gamma$ (magenta stars), $\gamma Z$ (green circles), and $\gamma h$ (blue squares) are shown for the leading (downward bars) and second (upward bars, bottom panel) triads.
        Energy ranges (bar widths) reflect likelihood bounds and not Gaussian uncertainties; the search window spans $3$--$100\GeV$ (unshaded).
    }
\end{figure}

The detrended X-ray-correlated spectrum, rebinned for visibility, is presented in Fig.~\ref{fig:Spectra} (filled black diamonds), alongside its ambient control counterpart $\delta I_0$ (empty red diamonds). While the former shows noticeable spectral features, the latter is consistent with noise.
Matched-filter features (colored bars) are extracted at the high native spectral resolution,
and are summarized in Table \ref{tab:Results}.

Error bars are well calibrated: over $3$--$100\GeV$, the normalized residuals $y_k/\sigma_k$ of $\delta I_\theta$ have a standard deviation of $1.03$, with extreme values ($-2.3$, $+3.0$), and those of the ambient control spectrum $\delta I_0$ have a standard deviation of $1.04$, with extreme values ($-3.0$, $+3.3$).
The residuals of $\delta I_0$ are symmetric, with four bins above $+2$ and four below $-2$,
whereas $\delta I_\theta$ shows seven bins above $+2$ and one below $-2$ (see \S\ref{subsec:Blind}).

\begin{table*}[t]
    \centering
    \caption{Spectral features in the X-ray-correlated spectrum $\delta I_\theta$.\label{tab:Results}}
    \setlength{\tabcolsep}{7pt}
    \begin{tabular}{l c c c c c c}
        \toprule
        Feature & $\epsilon$ & $Z_l$ & $Z_g\,(Z_g^{\smash[t]{(a)}})\,[Z_g^{\smash[t]{(G)}}]$ & $\{\epsilon_{\gamma\gamma},\epsilon_{\gamma Z},\epsilon_{\gamma h}\}$ & $\{Z_{\gamma\gamma},Z_{\gamma Z},Z_{\gamma h}\}$ & $\{\overline{\sigma v}_{p,\gamma\gamma},\overline{\sigma v}_{p,\gamma Z},\overline{\sigma v}_{p,\gamma h}\}/D$ \\
         & (GeV) & & & (GeV) & & ($10^{-19}\cm^3\se^{-1}$) \\
        \midrule
        \multirow{6}{*}{\shortstack[l]{Lines\\($f=0.1$)}} & $69.4_{-1.9}^{+1.5}$ & $2.6$ & $0.7\,(1.5)\,[0.8]$ & & & \\
         & $23.3_{-0.5}^{+0.6}$ & $2.1$ & & & & \\
         & $43.0_{-0.9}^{+0.8}$ & $2.0$ & & & & \\
         & $17.5_{-0.6}^{+0.5}$ & $1.5$ & & & & \\
         & $5.2_{-0.2}^{+0.3}$ & $1.5$ & & & & \\
         & $9.5_{-0.3}^{+0.2}$ & $1.4$ & & & & \\
        \midrule
        \multirow{4}{*}{\shortstack[l]{Triads\\($f=0.015$)}} & $71.9_{-0.3}^{+0.2}$ & $3.0$ & $1.8\,(2.1)\,[1.8]$ & $\{71.9,\,43.0,\,17.5\}$ & $\{2.0,\,1.7,\,1.5\}$ & $\{1.7\pm0.9,\,5.1\pm2.9,\,7.3\pm5.0\}$ \\
         & $67.5_{-0.2}^{+0.1}$ & $3.0$ & $1.7\,(2.0)\,[1.8]$ & $\{67.5,\,36.8,\,9.5\}$ & $\{2.4,\,1.4,\,1.4\}$ & $\{2.2\pm0.9,\,4.2\pm2.9,\,9.4\pm6.9\}$ \\
         & $87.7_{-0.8}^{+0.6}$ & $2.3$ & $0.4\,(1.0)\,[0.9]$ & $\{87.7,\,64.0,\,43.0\}$ & $\{0.3,\,1.7,\,2.0\}$ & $\{0.3\pm1.2,\,5.2\pm3.1,\,7.8\pm3.9\}$ \\
         & $65.3_{-0.1}^{+0.3}$ & $2.2$ & $0.2\,(0.9)\,[0.8]$ & $\{65.3,\,33.4,\,5.3\}$ & $\{1.8,\,0.6,\,1.4\}$ & $\{1.5\pm0.8,\,1.8\pm3.0,\,14.6\pm10.3\}$ \\
        \midrule
        \multirow{2}{*}{\shortstack[l]{Nonad\\($f=0.05$)}} & $66.4_{-1.4}^{+0.3}$ & \multirow{2}{*}{$3.2$} & \multirow{2}{*}{$2.5\,(2.5)\,[2.2]$} & & & \\
         & $71.9_{-0.6}^{+0.5}$ & & & & & \\
        \bottomrule
    \end{tabular}
    \TblNote{
        Matched-filter features: all $Z_l>1$ lines ($n_G=4.2\pm1.6$ expected in Gaussian noise; $n_G=2.3\pm1.4$ for $Z_l>1.5$) and $Z_l>2$ triads ($n_G=0.5\pm1.0$) and nonads ($n_G=0.2\pm0.5$). Expected numbers $n_G$ (mean and standard deviation) are from a separate set of $5\times10^4$ Gaussian noise spectra with the measured $\sigma_k$, scanned and selected in the same way. Features of different kinds, and triads sharing channels, are not independent; four or more $Z_l>2$ triads arise in $2.7\%$ of the noise spectra ($\smash{Z_g^{(G)}}\simeq1.9$).
        Energies and masses carry asymmetric $\Delta\textrm{TS}=1$ profile-likelihood bounds (not Gaussian uncertainties). Global $Z$-scores: nominal shuffle-calibrated $Z_g$, analytic \smash{\v{S}id\'{a}k} $\smash{Z_g^{(a)}}$ in parentheses,
        and Gaussian Monte Carlo $\smash{Z_g^{(G)}}$ in brackets, given for the strongest feature of each kind and for all triads.
        The $\epsilon$ column indicates $m_1c^2$ and $m_2c^2$ for the nonad (scanned with $f_t=0.02$) and $\overline{m}_\chi c^2$ for the triads, whose channel energies $\{\epsilon_{\gamma\gamma},\epsilon_{\gamma Z},\epsilon_{\gamma h}\}$ follow from Eq.~\eqref{eq:kin}.
        Cross sections are as measured; loss-corrected values require division by the appropriate $\mathcal{F}_i$ and $\mathcal{F}_{\mbox{\tiny PSF},i}$ factors (\S\ref{subsec:Losses}).
}
\end{table*}

\subsection{Blind scans}
\label{subsec:Blind}

The strongest single line ($f=0.1$) lies at $69.4_{-1.9}^{+1.5}\GeV$, with $Z_l\simeq2.6$.
Weaker, $Z_l\simeq2$ lines are picked up at $\sim23$ and $\sim43\GeV$.
The strongest line lies where the $\gamma\gamma$ channels of the two leading triads ($67.5$ and $71.9\GeV$; see below) blend, and where the $\gamma\gamma$ line of a $\sim72\GeV$ triad would appear if its emission were spread over $\mathcal{Z}(z)$
(which shifts measured single-line energies $\sim5$--$6\%$ below the emitted energy; \S\ref{subsec:Losses}).
The $\sim43\GeV$ line coincides with the $\gamma Z$ channel of the $71.9\GeV$ triad (\S\ref{sec:43GeV}), and the $\sim23\GeV$ line with the $\gamma h$ channel of a $\sim75.6\GeV$ triad, at the upper mass of the leading nonad of $\delta I$ (Appendix \ref{app:dI}).
The three strongest lines of $\delta I$ similarly coincide with triad channels (Appendix \ref{app:dI}).

The most significant triad lies at $\overline{m}_\chi c^2=71.9_{-0.3}^{+0.2}\GeV$, with $Z_l\simeq3.0$ and all three channels contributing comparably.
Note that the wide, $\sim10$--$70\GeV$ dynamic range of such kinematically fixed triads acts as a built-in veto against localized instrumental or background artifacts.
The global significance of the leading triad is $Z_g\simeq1.8$ ($2\times10^4$ shuffled-contrast skies; $p_g\simeq0.04$), with $\smash{Z_g^{(G)}}\simeq1.8$ (median of highest simulated $Z_l$ is $1.6$), $\smash{Z_g^{(n)}}\simeq1.9$, and $\smash{Z_g^{(a)}}\simeq2.1$.
A second triad, of nearly equal significance, lies at $67.5_{-0.2}^{+0.1}\GeV$ ($Z_l\simeq3.0$, $Z_g\simeq1.7$); other triads have $Z_l<2.5$.

The two leading triads share the broad excess near $\sim70\GeV$ in their $\gamma\gamma$ channels, which lie $6\%$ apart, within the core of the IRF energy dispersion.
Their other channels differ: $\sim43.0$ and $\sim17.5\GeV$ for the $71.9\GeV$ triad, as opposed to $\sim36.8$ and $\sim9.5\GeV$ for the $67.5\GeV$ triad.
The measured brightnesses (before correcting for pipeline and PSF losses; \S\ref{subsec:Losses}) correspond (for the two triads) to $\overline{\sigma v}_{p,i}/D\simeq(1.7$--$2.2)$, $(4.2$--$5.1)$, and $(7$--$9)\times10^{-19}\cm^3\se^{-1}$ cross sections in the $\gamma\gamma$, $\gamma Z$, and $\gamma h$ channels, respectively.
These values (Table \ref{tab:Results}) suggest a crude $\gamma\gamma<\gamma Z<\gamma h$ hierarchy, albeit with large uncertainties.

The most significant nonad, with $m_1c^2\simeq66.4$ and $m_2c^2\simeq71.9\GeV$, has $Z_l\simeq3.2$ and a global $Z_g\simeq2.5$ ($\smash{Z_g^{(G)}}\simeq2.2$); as its outer masses nearly coincide with those of the two leading triads, it largely combines them.
The intermediate triad of this nonad, at $\sim69.2\GeV$,
adds a conditional $Z_c\simeq0.4$ score beyond the correlated contribution of the two outer triads, a value typical of
equal-brightness mock nonads at such masses (see Appendix~\ref{app:Mock}).
A second nonad maximum, with $Z_l\simeq3.16$ and $Z_g\simeq2.4$ (against $3.19$ for the leading nonad) at $\{m_1,m_2\}c^2\simeq\{67.4,75.6\}\GeV$, ranks first under equal-$Z$ weighting and in $\delta I$.

The LSS-enhanced sum $\delta I$ recovers the same features, with the expected higher noise (Appendix \ref{app:dI}): the strongest line, at $69.3\GeV$, and the two leading triads,
with the $67.5\GeV$ one ($Z_l\simeq3.3$, global $Z_g\simeq1.7$; $\smash{Z_g^{(G)}}\simeq2.2$) slightly stronger than in $\delta I_\theta$ and the $71.8\GeV$ one ($Z_l\simeq2.8$) slightly weaker.
The leading nonad of $\delta I$, at $\{67.4,75.6\}\GeV$ ($Z_l\simeq3.3$, global $Z_g\simeq1.8$; $\smash{Z_g^{(G)}}\simeq2.4$), contains the same two triads, here as its lower and intermediate ($\sim71.5\GeV$) triads.

Thus, the blind scans provide a global $\sim2\sigma$ hint of $\overline{m}_\chi c^2\simeq71.9$ and $67.5\GeV$ triads, or of a nonad combining them with a weaker intermediate triad, insufficient on its own for a detection; however, see \S\ref{sec:43GeV}.
The mock triads show measured $Z_l$ typically lower than injected, by $\sim10$--$30\%$ (Appendix \ref{app:Mock}), indicating that detrending reduces the sensitivity to a real signal.
Nonads lose more: at the injected masses, the measured $Z_l$ of $\{66.4,71.9\}\GeV$ mock nonads falls $\sim30\%$ below the injected value, against $\sim15\%$ for $71.9\GeV$ mock triads injected with the same $Z_l$ into the same skies (Appendix~\ref{app:Mock}).
Hence, the measured $Z_l$ of the $71.9$ and $67.5\GeV$ triads and of the $\{66.4,71.9\}\GeV$ nonad
typically underestimates that of a real signal.

A test without matched filters examines if the spectrum is consistent with emission lines, which add only positive residuals.
We thus compare the numbers of bins with normalized $y_k/\sigma_k>2$ vs. $y_k/\sigma_k<-2$ residuals; the difference is calibrated by $2\times10^4$ shuffled-contrast skies (Appendix \ref{app:MC}).
The $\delta I_\theta$ spectrum has six more positive than negative such bins (out of $170$ bins within $3$--$100\GeV$), a difference matched or exceeded by only $0.47\%$ of the shuffled-contrast skies ($2.6\sigma$).
Restricting the count to $3$--$80\GeV$, away from photon starvation at high energies, gives $2.5\sigma$.
This excess arises largely in the bins of triad channels, so while consistent with an annihilation signal it is not independent of triad and nonad results.

Other asymmetry statistics, calibrated by the same shuffled-contrast skies, are less significant, with $p$-values of $0.01$ for the normalized signed sum of $\smash{y_k^2/\sigma_k^2}$, and $0.02$ for the skewness of $y_k/\sigma_k$.
A conservative \smash{\v{S}id\'{a}k} correction for the three asymmetry statistics gives $p\simeq0.014$ ($2.2\sigma$) for the excess of positive bins; the $3$--$80\GeV$ range gives $2.1\sigma$.

\subsection{Redshift-broadened apertures}
\label{subsec:Dz}

\begin{figure}[t!]
    \includegraphics[width=\columnwidth]{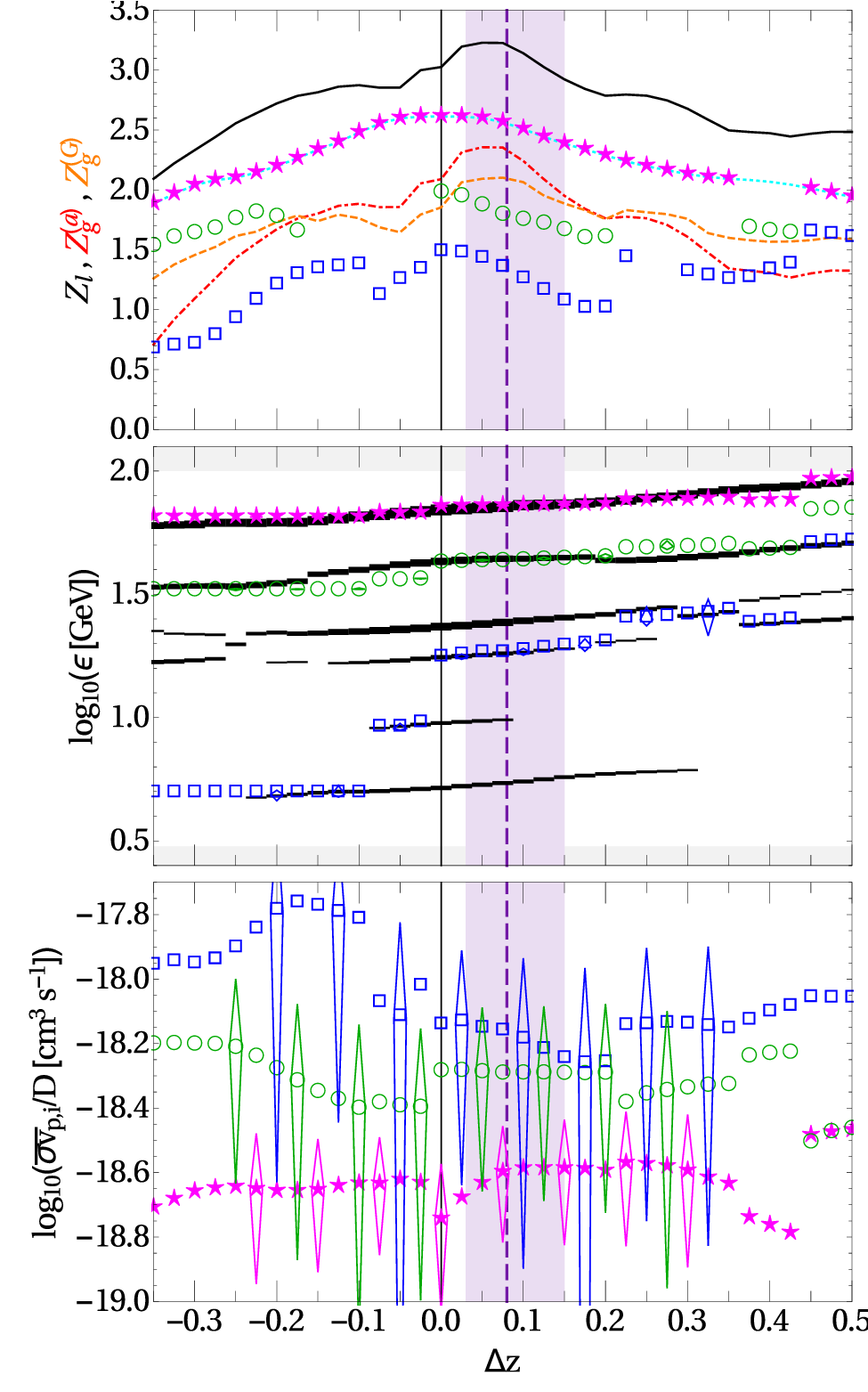}
	\caption{\label{fig:Dz}
        Response of the $\delta I_\theta$ triad filter to redshift-broadening the aperture from $z=0$ to $\Delta z$.
        The peak position expected for $\mathcal{Z}(z)$ is marked in all panels (median of the mock triads of \S\ref{subsec:Dz}, dashed purple; their $16$--$84\%$ range, purple band).
        Top: local $Z_l$ of the strongest triad at each $\Delta z$ (solid black), its global (but not accounting for the $\Delta z$ scan) $\smash{Z_g^{(a)}}$ (dot-dashed red) and $\smash{Z_g^{(G)}}$ (dashed orange; Monte Carlo at each fixed $\Delta z$), the largest single-line $Z_l$ (dotted cyan), and the $Z_l$ of the lines nearest to the triad channels (symbols).
        Middle: triad $\gamma\gamma$ (magenta stars), $\gamma Z$ (green circles), and $\gamma h$ (blue squares) channel energies, with $\Delta\textrm{TS}=1$ bounds (diamonds, for $Z_l>2.5$, at every third $\Delta z$ value), and single lines with $Z_l>1$ (black bars of thickness rising linearly with $Z_l$), within the $3$--$100\GeV$ search window (unshaded).
        Bottom: measured intrinsic channel cross sections (symbols as above).
    }
\end{figure}

Figure \ref{fig:Dz} follows the strongest $\delta I_\theta$ triad at each $\Delta z$, as the triad aperture is broadened by Eq.~\eqref{eq:Dz} from $z=0$ to $\Delta z$, showing its significance (top panel), the energies of its three channels (middle), and their cross sections (bottom), together with the single lines picked up at each $\Delta z$ (top and middle).
Positive $\Delta z$ spreads each channel towards lower energies, as cosmological redshifts do, whereas negative $\Delta z$ values, unphysical for cosmological redshifts, spread it towards higher energies, and so probe the spectral structure just above the triad channels.

A signal emitted at $z=0$, such as an instrumental or Galactic feature, would peak at $\Delta z=0$, whereas a signal spread over low redshifts would peak at a positive $\Delta z$ of the order of its redshift spread; in both cases, $Z_l$ declines once the broadening exceeds the signal spread and dephases the rigid kinematic structure of the triad.

The leading triad, of $\overline{m}_\chi c^2\simeq71.9\GeV$ at $\Delta z=0$, strengthens with positive broadening, from $Z_l\simeq3.03$ to a maximum $Z_l\simeq3.25$ at $\Delta z\simeq0.05$ ($\overline{m}_\chi c^2\simeq72.2\GeV$), and then declines.
This preferred broadening corresponds to a mean redshift $\Delta z/2\simeq0.025$, close to the redshift at which the anticipated LSS signal peaks ($\sim0.03$; Appendix \ref{app:PeakZ}), consistent with an annihilation signal dominated by low-redshift structure.
The $0.22$ gain in $Z_l$ is not adopted as nominal, as $\Delta z$ is an additional scanned parameter, not included in the trial correction.
With the coarser ($s_0\simeq3.7\%$) native binning of Appendix \ref{app:Nominal}, this triad similarly strengthens, from $Z_l\simeq2.5$ to $2.6$ at $\Delta z\simeq0.05$.

To verify the viability of this test, we inject mock triads, with equal channel-$Z_l$ values and spread over $\mathcal{Z}(z)$, into the correlated spectra of shuffled-contrast skies, before detrending.
Their brightness is scaled so that the injected signal alone, passed through the full pipeline and scored with the errors $\sigma_k$ of the data, gives $Z_l\simeq3$ at $\Delta z=0$, as observed.
These mocks show a similar $Z_l$ gain (median $0.26$) and peak position (median $\Delta z\simeq0.08$; $16$--$84\%$ range $0.03$--$0.15$).
Similarly scaled $z=0$ mock triads do not show such an effect (gain exceeding $0.22$ with a $\Delta z\gtrsim0.05$ peak in only $\sim3\%$ of cases).

The physically relevant range of $\Delta z$ is narrow; the exaggerated range in the figure is intended to pick up neighboring triads.
Indeed, the discontinuities in the channel energies (middle panel), and the corresponding steps in the other panels, mark changes in the identity of the leading triad, as the scan follows the strongest triad at each $\Delta z$.
For negative $\Delta z$, the strongest triad shifts to $\overline{m}_\chi c^2\simeq67.3$--$67.5\GeV$ and then to $65.1\GeV$, reflecting the second and fourth triads of Table \ref{tab:Results}.

At $\Delta z=0$, the triad channels coincide, within the energy resolution, with single lines of the spectrum (middle panel).
The $\gamma\gamma$ channel thus agrees with the strongest, $\sim69\GeV$ line, the $\gamma Z$ channel with the $\sim43\GeV$ line, and the $\gamma h$ channel with the $\sim17.5\GeV$ line.
The inferred cross sections (bottom panel) show the crude $\gamma\gamma<\gamma Z<\gamma h$ hierarchy of \S\ref{subsec:Blind} throughout the scanned range of $\Delta z$, albeit with large uncertainties.

\subsection{Robustness}
\label{subsec:Robustness}

Instrumental or sky artifacts would not naturally produce the composite spectral features observed in $\delta I_\theta$, or bias its residuals positively.
Such artifacts would also not reproduce, in a blind scan, the previously reported line $L_{43}$, prefer positive over negative redshift broadening of the aperture, and peak at $\Delta z\simeq 0.05$, as discussed above. Additional properties that artifacts would not generally present are outlined below.

By design, artifacts are minimized by choosing the cleanest \gama-ray data (highest purity event class, most restrictive IRF), LSS tracers (minimal NXB in eROSITA maps), and pipelines (recommended Pass 8 cuts, aggressive masking of contaminations).
Tests of alternative eROSITA bands, NXB-subtracted maps, and varied ROI cuts showed no significant changes in the features, which accumulated similarly in different LAT epochs.
These comparative tests use the pseudo-count Poisson errors of Appendix \ref{app:dI}, which, by omitting the X-ray-correlated variance, make any difference between the compared spectra appear more significant.

The ambient spectrum $\delta I_0$ provides a powerful check on a wide range of artifacts.
Although the ambient spectrum is $\smash{\llangle\boldsymbol{X}_\theta^2\rrangle^{-1/2}}\simeq2.3$ times more sensitive than the X-ray-correlated spectrum, it shows none of the $\delta I_\theta$ spectral features.
Scanned and selected as in Table~\ref{tab:Results}, $\delta I_0$ shows no $Z_l>2$ triads or nonads; its strongest triad ($Z_l\simeq1.7$; $\smash{Z_g^{(G)}}\simeq0.1$) and nonad ($Z_l\simeq1.7$; $\smash{Z_g^{(G)}}\simeq0.6$) are typical of Gaussian noise.
At the specific energies of all features of Table~\ref{tab:Results} and of the externally fixed line, triad, and nonad of \S\ref{sec:43GeV}, the ambient spectrum is consistent with zero ($|Z_l|\leq1.1$); in particular, $Z_l\simeq-1.1$ at $43.2\GeV$, $(-0.85)$ for the nominal triad, and $0.35$ for the blind nonad.

Any instrumental, event-reconstruction, sky, or analysis artifact acting on \gama-rays independently of the X-ray contrast enters $\delta I_\theta$ only through the correlated subset of \gama-rays.
If produced by such artifacts, the $\delta I_\theta$ features would therefore be brighter in $\delta I_0$ by the $\sim25$--$130$ ratio of the ambient to X-ray-correlated continuum, and thus reach $Z_l\simeq130$--$460$.
As $\delta I_0$ limits each spectral feature to at most $22\%$--$63\%$ of its $\delta I_\theta$ brightness
($95\%$ confidence), rather than the $\sim25$--$130$ times expected for an artifact, such artifacts account, feature by feature, for $\lesssim1.3\%$ of any reported $Z_l$.
The $\delta I_\theta$ features are thus securely attributed to X-ray-correlated emission, with $\delta I_0$ showing nothing beyond the diluted sky mean of this component.

The results are insensitive to the detrending scale. For instance, varying the GPR correlation length between $0.18$ and $0.63$ decades in energy changes the $Z_l$ of the two leading triads by $\leq0.03$, and the nonad by $\leq 0.04$.
As a robustness test, applying the matched filters to spectra rebinned by grouping pairs of native bins ($s\simeq2s_0\simeq4.2\%$) changes these $Z_l$ values by $\lesssim0.2$, whereas a coarser four-bin grouping ($s\simeq4s_0\simeq8.4\%$), with bins comparable to the $\sim9\%$ full width at half maximum of the compact IRF core, introduces fluctuations of up to $\sim0.7$ in $Z_l$.
As a consistency check, synthetic apertures (symmetric and asymmetric top-hats and Crystal Ball functions) are found to recover features at energies very similar to those found with the IRF, and with expectedly diminished $Z_l$.

When the analyzed sky is split into independent halves, the nominal triad appears both north ($Z_l\simeq2.3$) and south ($Z_l\simeq2.8$) of the Galactic plane and in both longitude halves (split at $l\simeq245\dgr$, the median longitude of the analyzed pixels, so that the halves have equal pixel counts), consistent with a signal distributed across the sky.
The $l>245\dgr$ half gives $Z_l\simeq3.5$, against $1.2$ for $l<245\dgr$ (differing by $1.9$ standard deviations); the former harbors Virgo, the Horologium--Reticulum supercluster, and most of the supergalactic plane within the analyzed sky.
The blind scan of the stronger longitude half alone recovers both leading triads, at $72.0$ and $67.5\GeV$ ($Z_l\simeq3.5$ and $3.3$).
The blind scan of the southern half finds its strongest triad at $87.9\GeV$, with its $\gamma h$ channel coinciding with the $43\GeV$ line, which is strong in this southern region.

Masking the Virgo and Fornax clusters, both within the analyzed sky, leaves the results nearly unchanged (\S\ref{sec:43GeV}).
The Magellanic Clouds also lie within the analyzed sky.
Excluding the pixels within $10\dgr$ of the Large Magellanic Cloud or of the nearby south ecliptic pole, or within $5\dgr$ of the Small Magellanic Cloud ($4\%$ of the pixels altogether), changes the nominal triad only from $Z_l\simeq3.0$ to $2.8$, consistent with the reduced statistics, and leaves the leading triad at $71.9\GeV$.

The $[-5,+2]\%$ absolute systematic uncertainty in the LAT energy scale \cite{AckermannEtAl12EnergyScale, FermiCalibration12} is the dominant source of systematic error in the inferred line energies and WIMP masses.
A scale error would shift all observed channel energies by the same fraction, misaligning them with the triad template, whose relative channel energies are non-linearly fixed by $m_Z$ and $m_h$ in Eq.~\eqref{eq:kin}, and would also bias the inferred brightnesses and cross sections.
Current uncertainties in $m_Z$ and $m_h$ are sufficiently small to have a negligible effect on the results.

\section{Implications of the $\sim43\GeV$ cluster line}
\label{sec:43GeV}

Interpreting $L_{43}$, the $\epsilon\simeq43.2\GeV$ line reported in the Virgo, Fornax, and Ophiuchus clusters \cite{FanEtAl26}, as a WIMP annihilation channel, Eq.~\eqref{eq:kin} fixes $\overline{m}_\chi c^2$ at $\sim72.1\GeV$ for the $\gamma Z$ channel or $\sim87.8\GeV$ for the $\gamma h$ channel; an alternative, $\gamma\gamma$ reading would give $\overline{m}_\chi c^2\simeq43.2\GeV$, below the triad range.
Remarkably, the $\gamma Z$ reading coincides with the most significant triad, at $71.9_{-0.3}^{+0.2}\GeV$, whereas the $\gamma h$ reading coincides with the third triad, at $87.7_{-0.8}^{+0.6}\GeV$ (see Table \ref{tab:Results}).
At these externally fixed masses, no mass scan is involved, so there is no need for a multiple-trial correction as for blind scans.
The present and three-cluster LAT analyses share the same energy scale,
so a LAT scale error would shift $L_{43}$ and the $\sim43\GeV$ line of the present data alike.

The apertures are centered at $z=0$, whereas the signal is emitted over a range of redshifts, so the mass we infer from the observed lines is somewhat lower than the true WIMP mass.
Mock triads injected with the channel brightness ratios measured for each of the two leading triads (in both, the $\gamma\gamma$ channel is the brightest), spread over $\mathcal{Z}(z)$ and passed through the full pipeline, are inferred a median $\sim0.9\%$ below their true mass.
This offset is far smaller than the $\sim5$--$6\%$ shift of single lines (\S\ref{subsec:Losses}), because a given redshift lowers the masses implied by the $\gamma Z$ and $\gamma h$ lines only $\sim0.43$ and $\sim0.14$ times as much as that implied by the $\gamma\gamma$ line (Eq.~\eqref{eq:kin}),
so the $\gamma Z$ and $\gamma h$ lines pin the fitted mass.

The offset depends, however, on the uncertain measured ratios, and would approach the single-line shift for a $\gamma\gamma$-dominated triad.
Correcting for the $\sim0.9\%$ offset would place the $71.9\GeV$ triad at $\sim72.6\GeV$.
An aperture broadened to the preferred $\Delta z\simeq0.05$ instead places it at $\sim72.2\GeV$ (see \S\ref{subsec:Dz}).
Both values lie within $\sim0.8\%$ of the $72.1\GeV$ mass of the $\gamma Z$ reading.
The $\pm0.5\GeV$ uncertainty of the three-cluster $L_{43}$ energy ($43.7\pm0.6\GeV$ in a 13-cluster sample\cite{FanEtAl26})
corresponds to $\pm0.4\GeV$ in the $\gamma Z$-reading mass, consistent with the above $71.9$--$72.6\GeV$ estimates; the fixed-mass tests below use the central $72.1\GeV$ mass.

Table~\ref{tab:Fixed} summarizes the tests at these externally fixed parameters for three features: the line at the $L_{43}$ energy, the nominal triad, and the nonad with one mass fixed at $72.1\GeV$ and the other scanned.
For each feature, the table lists the local $Z_l$, also with Virgo and Fornax masked and in the ambient spectrum $\delta I_0$, followed by its calibrated significance as a single test ($Z_g$ from shuffled-contrast skies, $\smash{Z_g^{(G)}}$ from the Gaussian calibration) and the combination of $Z_g$ with $L_{43}$ or $L_{43b}$.
The lower block repeats the calibrations when allowing either the $\gamma Z$ or the $\gamma h$ reading.
Parentheses give the two companion (non-$43\GeV$) channels of the triad, and brackets give the leading blind triad and nonad of Table~\ref{tab:Results}, which lie at nearly the same masses but carry the multiple-trial correction of their scans.

\begin{table}[t]
    \centering
    \caption{X-ray-correlated features at externally fixed mass.
    \label{tab:Fixed}}
    \setlength{\tabcolsep}{4.5pt}
    \begin{tabular}{l c c c}
        \toprule
         & Line & Triad & Nonad \\
        \midrule
        $\epsilon$ or $mc^2$ (GeV) & $43.2$ & $72.1$ & $\{65.4,72.1\}$ \\
         & & $[71.9]$ & $[\{66.4,71.9\}]$ \\
        $Z_l$ & $1.97$ & $2.97\,[3.03]$ & $3.16\,[3.19]$ \\
        $Z_l$, masked & $2.07$ & $2.91\,[2.98]$ & $3.15\,[3.19]$ \\
        $Z_l$, ambient $\delta I_0$ & $-1.05$ & $-0.85\,[-0.67]$ & $-0.02\,[0.35]$ \\
        $Z_g$ & $1.7$ & $3.0\,(2.4)\,[1.8]$ & $2.9\,[2.5]$ \\
        $Z_g^{(G)}$ & $2.0$ & $3.0\,(2.4)\,[1.8]$ & $2.6\,[2.2]$ \\
        $Z_g$ with $L_{43}$ & $4.4$ & $5.1\,(4.7)\,[4.4]$ & $5.0\,[4.8]$ \\
        $Z_g$ with $L_{43b}$ & $3.0$ & $3.8\,(3.4)\,[3.0]$ & $3.8\,[3.5]$ \\
        \midrule
        \multicolumn{4}{l}{Allowing either reading ($\gamma Z$ or $\gamma h$)} \\
        $Z_g$ & --- & $2.9\,(2.2)$ & $2.9$ \\
        $Z_g^{(G)}$ & --- & $2.7\,(2.1)$ & $2.5$ \\
        $Z_g$ with $L_{43}$ & --- & $5.0\,(4.6)$ & $5.0$ \\
        $Z_g$ with $L_{43b}$ & --- & $3.8\,(3.3)$ & $3.8$ \\
        \bottomrule
    \end{tabular}
    \TblNote{
    Fixed features: the line at the $L_{43}$ energy, the nominal triad, and the nonad with one mass fixed at $72.1\GeV$ and the other scanned (largest interior maximum). The lower block allows either reading (also $87.8\GeV$ for $\gamma h$; the larger statistic). Parentheses: the two companion (non-$43\GeV$) channels of the triad. Brackets: the leading blind triad and nonad (Table~\ref{tab:Results}), at their own masses. Masked: Virgo and Fornax masked within the radii of Ref.~\citenum{FanEtAl26}. Values of $Z_g$ are calibrated by $2\times10^4$ shuffled-contrast skies, and $\smash{Z_g^{(G)}}$ by Gaussian Monte Carlo. Combinations with $L_{43}$ ($4.3\sigma$) and $L_{43b}$ ($2.7\sigma$) use Fisher's method\cite{Fisher25} with the shuffle-calibrated $p$-values.}
\end{table}

As shown in Table \ref{tab:Fixed}, the nominal triad reaches $Z_l\simeq3.0$, with all three of its channels contributing: $\{Z_{\gamma\gamma},Z_{\gamma Z},Z_{\gamma h}\}\simeq\{2.0,1.7,1.4\}$.
Shuffled-contrast skies calibrate this as a $p$-value of $1.3\times10^{-3}$ ($3.0\sigma$) for the $\gamma Z$ reading, or $1.6\times10^{-3}$ ($2.9\sigma$)
when also allowing the $\gamma h$ reading
(Appendix~\ref{app:MC}),
in agreement with the Gaussian calibration (Table~\ref{tab:Fixed}).

As in the case of blind triads (\S\ref{subsec:Blind}), measured $Z_l$
typically underestimates that of a real signal: at the injected mass, the measured $Z_l$ of $71.9\GeV$ mock triads falls $\sim15\%$ below the injected value (Appendix~\ref{app:Mock}).
Accounting for the redshift of the signal, the nominal triad gives $Z_l\simeq3.2$ ($3.2\sigma$, not corrected for the choice of $\Delta z$) with an aperture broadened to $\Delta z\simeq0.05$ (see \S\ref{subsec:Dz}), $Z_l\simeq2.4$ ($2.3\sigma$) when instead shifting the aperture down by the $\sim0.9\%$ offset of the specific mocks above, and $Z_l\simeq2.5$ ($2.3\sigma$) when spreading it over $\mathcal{Z}(z)$.

It is interesting to examine the nominal triad without matched filters, simply adding the pulls $y_k/\sigma_k$ of the bins holding the three channel energies, rebinned as in Fig.~\ref{fig:Spectra} and averaged over the bin-grid phase.
This simple approach gives $3.1\sigma$ ($2.9\sigma$) at the medium (low) resolution, as calibrated by $2\times10^4$ shuffled-contrast skies.
Relative to the same, negative sum in $\delta I_0$, with its noise taken as Gaussian, these become $3.7\sigma$ ($3.2\sigma$).

The present and three-cluster analyses are practically independent.
Virgo and Fornax lie within the analyzed western sky, but masking them within the radii adopted in Ref.~\citenum{FanEtAl26} (removing $0.8\%$ of the analyzed pixels) changes the nominal triad only from $Z_l\simeq3.0$ to $2.9$, and masking twice these radii ($3.5\%$ of the pixels) lowers it to $2.8$; both results are consistent with the reduced statistics.
Ophiuchus lies outside our analyzed sky.
Combining the $p$-values of the two analyses by Fisher's method\cite{Fisher25}
yields $5.1\sigma$ ($3.8\sigma$) with $L_{43}$ ($L_{43b}$), or $5.0\sigma$ ($3.8\sigma$) when also allowing the $\gamma h$ reading.
The leading blind triad of Table~\ref{tab:Results}, with its multiple-trial correction, gives $4.4\sigma$ ($3.0\sigma$) with $L_{43}$ ($L_{43b}$).
Over the above redshift treatments, the $\gamma Z$ reading combined with $L_{43}$ ($L_{43b}$) gives $4.7$--$5.2\sigma$ ($3.3$--$4.0\sigma$).

The nominal triad implies the intrinsic primary-channel ($\gamma\gamma$) cross section, corrected for pipeline and PSF losses (\S\ref{subsec:Losses}),
\begin{equation}\label{eq:Sigma0}
  \Sigma_0 \equiv \frac{\overline{\sigma v}_{p,\gamma\gamma}}{D} = (10.8 \pm 8.7) \times 10^{-19} \cm^3 \se^{-1} \, ,
\end{equation}
for $\mathcal{Z}(z)$, and $\overline{\sigma v}_{p,i}/(10^{-19}D\cm^3\se^{-1}) = (28\pm29)$ and $(46\pm34)$ in the $\gamma Z$ and $\gamma h$ channels.
As about a third of the modeled signal, emitted at $z\gtrsim0.1$, is shifted out of the $z=0$ apertures and essentially not recovered (Appendix \ref{app:PeakZ}), $\Sigma_0$ depends on $\mathcal{Z}(z)$.
Truncating $\mathcal{Z}(z)$ at $z=0.1$ gives $\Sigma_0\simeq4.7\times10^{-19}\cm^3\se^{-1}$, a uniform distribution over $z<0.05$ gives $3.2\times10^{-19}\cm^3\se^{-1}$, and correcting only for detrending losses ($\mathcal{F}_i=\mathcal{F}_{0,i}$ and $\mathcal{F}_{\mbox{\tiny PSF},i}=1$)
gives $\Sigma_0\simeq2.2\times10^{-19}\cm^3\se^{-1}$, a lower limit for any redshift distribution.
All these cross sections assume $\mathcal{F}_b=1$, so, by the bound $\mathcal{F}_b\lesssim0.6$ of Appendix~\ref{app:LossFactors}, each is underestimated by a factor $\gtrsim1.6$.

The present data also show a line at the $L_{43}$ energy, independently of any triad interpretation.
At the externally fixed $43.2\GeV$ energy, the single-line $Z_l\simeq2.0$ (Table \ref{tab:Fixed}) is calibrated by shuffled-contrast skies as a $p$-value of $0.05$ ($1.7\sigma$).
Combined with $L_{43}$ ($L_{43b}$), this line gives $4.4\sigma$ ($3.0\sigma$).
Masking Virgo and Fornax within the radii of Ref.~\citenum{FanEtAl26} (twice these radii) does not weaken this feature: its $Z_l$ rises slightly, from $1.97$ to $2.07$ (to $2.05$),
consistent with the spatially extended emission reported \cite{Profumo26} and expected for $p$-wave annihilation.

One might suspect that the X-ray-correlated triads could merely reflect a $43\GeV$ line, with no accompanying channels.
Such a line alone could produce triads at both $72.1$ and $87.8\GeV$, as found here (Table \ref{tab:Results}), but with only noise in their other channels.
In either reading, the triad would then reach only $Z_l\simeq Z_{43}/\smash{\sqrt{3}}\simeq1$,
where $Z_{43}\simeq1.7$--$2.0$ is the $Z$-score of its $43\GeV$ channel.
For the $\gamma Z$ reading, this is well below the measured value: the observed $\gamma\gamma$ and $\gamma h$ channels alone give $(Z_{\gamma\gamma}+Z_{\gamma h})/\smash{\sqrt{2}}\simeq2.4$, calibrated by shuffled-contrast skies as a $p$-value of $8.2\times10^{-3}$ ($2.4\sigma$).
In contrast, the $\gamma\gamma$ and $\gamma Z$ channels of the third triad, which matches the $\gamma h$ reading, give only $\sim1.4$; allowing for either reading thus gives $2.2\sigma$.

Hence, the present data support the $\gamma\gamma$ and $\gamma h$ lines predicted by the $\gamma Z$ reading at the $\sim2.4\sigma$ level, independently of any $43\GeV$ emission.
Combined with $L_{43}$ ($L_{43b}$), these two channels alone give $4.7\sigma$ ($3.4\sigma$).
The significance of the nominal triad is slightly lower in the noisier sum $\delta I$ ($2.8\sigma$; Appendix \ref{app:dI}) and lower with a coarser native binning ($2.3\sigma$ in the Gaussian calibration; Appendix \ref{app:Nominal}).

The leading nonad can similarly be tested at the externally fixed mass, with its second mass left free.
Fixing one nonad mass at $72.1\GeV$ and scanning the other finds a $\{m_1,m_2\}c^2\simeq\{65.4,72.1\}\GeV$ nonad at $Z_l\simeq3.2$, with its free mass bounded within the same $65.0$--$66.7\GeV$ range ($65.4_{-0.4}^{+1.3}\GeV$) as the lower mass of the blind nonad (Table~\ref{tab:Results}).
Shuffled-contrast skies calibrate this as $2.9\sigma$ (Appendix~\ref{app:MC}), comparable to the triad alone despite the free mass.
Masking Virgo and Fornax changes its $Z_l$ from $3.16$ to $3.15$ (or $2.90$ for twice the radii), similar to the triad, and the ambient spectrum shows $Z_l\simeq-0.02$ at its masses.
Combined with $L_{43}$ ($L_{43b}$), this nonad gives $5.0\sigma$ ($3.8\sigma$); the blind nonad, with its multiple-trial correction, gives $4.8\sigma$ ($3.5\sigma$).

The externally fixed mass may instead be the cross-annihilation (mean) mass of the two species.
Fixing the mean nonad mass at $72.1\GeV$ and scanning the lower mass finds a $\{m_1,m_2\}c^2\simeq\{68.4,75.7\}\GeV$ nonad at $Z_l\simeq3.1$,
near the second blind nonad maximum at $\{67.4,75.6\}\GeV$ (see \S\ref{subsec:Blind}).
Shuffled-contrast skies calibrate this nonad as $3.0\sigma$, and $5.1\sigma$ ($3.9\sigma$) when combined with $L_{43}$ ($L_{43b}$).
The difference between the cross- and self-annihilation $Z_l$ values is not significant enough to distinguish the two roles of the $72.1\GeV$ mass;
allowing either role would give $2.9\sigma$, or $5.0\sigma$ ($3.8\sigma$) if combined with $L_{43}$ ($L_{43b}$).

For a $\overline{m}_\chi c^2\simeq72.1\GeV$ WIMP, the clusters showing $L_{43}$
should also show a $\gamma\gamma$ line at $72.1\GeV$ and a $\gamma h$ line at $17.7\GeV$.
However, with only $\sim5$--$10$ photons per line per cluster, these companions are not yet testable.
The tension posed by the absence of $L_{43}$ in the inner Galaxy, noted in Ref.~\citenum{FanEtAl26} as disfavoring canonical annihilation, is partly alleviated for $p$-wave annihilation, which is suppressed there by the $\sim10^{2}$ times lower $\beta^2$ ($\sim10^{-7}$ vs. $\sim10^{-5}$ in clusters).

\section{Summary and discussion}
\label{sec:Summary}

We cross-correlate the high-latitude, high spectral resolution ($s_0\simeq2.1\%$) 16.3-year LAT data with the eROSITA map of the western Galactic hemisphere.
A blind matched-filter scan for WIMP annihilation-line features suggests spectral lines, triads ($\gamma\gamma$, $\gamma Z$, and $\gamma h$ lines from the same center-of-mass energy), and a nonad from two cross-annihilating WIMPs.
These features, evident in the rebinned correlated spectrum (filled black diamonds) of Fig.~\ref{fig:Spectra}, are absent in the control, ambient spectrum (empty red diamonds).

The leading triads are found with mean annihilating-WIMP masses $\overline{m}_\chi c^2\simeq71.9$ and $67.5\GeV$ at (global, henceforth) $1.8\sigma$ and $1.7\sigma$.
A nonad combining them and an additional, intermediate triad emerges at $2.5\sigma$.
These features, summarized in Table \ref{tab:Results} and Appendix \ref{app:dI}, are suggestive, but insufficiently strong on their own for an independent detection.

However, the $\gamma Z$ channel of the leading $\sim72\GeV$ triad coincides with the $\sim43.2\GeV$ line $L_{43}$ reported in three nearby clusters \cite{FanEtAl26}; our data show a $1.7\sigma$ line at this energy, not driven by these three clusters.
The $\gamma\gamma$ and $\gamma h$ channels of the nominal triad reach $\sim2.4\sigma$, independently of any $43\GeV$ emission, suggesting structure beyond this single line.
The $\sim72\GeV$ triad also strengthens when its aperture is broadened towards positive redshifts (peaking at $\Delta z\simeq0.05$), consistent with emission from low-redshift LSS (\S\ref{subsec:Dz}).

At this $\overline{m}_{\chi}c^2\simeq72.1\GeV$ mass, externally fixed by $L_{43}$, the nominal triad is significant at $3.0\sigma$ ($2.3$--$3.2\sigma$ for different treatments of the signal redshift; $2.9\sigma$
when also allowing the $\gamma h$ reading).
Combined with $L_{43}$ ($L_{43b}$), the triad reaches $5.1\sigma$ ($3.8\sigma$), and $4.7$--$5.2\sigma$ ($3.3$--$4.0\sigma$) depending on the redshift treatment (Table \ref{tab:Fixed} and \S\ref{sec:43GeV}).

The two leading triads suggest more than one annihilating WIMP species, so cross-annihilation must also be considered.
The nonad scan, which combines two self-annihilation triads with a cross-annihilation triad, finds $\{m_1,m_2\}c^2\simeq\{66.4,71.9\}\GeV$ at $2.5\sigma$.
The $\sim69.2\GeV$ cross-annihilation triad of this nonad adds a conditional $Z_c\simeq0.4$ score beyond its two self-annihilation triads, typical of mock nonads (\S\ref{subsec:Blind}), so the data are consistent with a genuine nonad.

Finding a nonad is notable, as the detrending suppresses nonads more than triads: the measured $Z_l$ of $\{66.4,71.9\}\GeV$ mock nonads falls $\sim30\%$ below the injected value, against $\sim15\%$ for $71.9\GeV$ triads, and nonads of random masses are recovered in only $\sim20$--$45\%$ of cases, against $\gtrsim90\%$ for triads (Appendix~\ref{app:Mock}).
The null skies offset this loss only in part: the detrending reduces the dispersion of their noise-driven $Z_l$ less than it reduces a real nonad, whose nine lines, crowded within the detrending scale, are partly absorbed into the continuum.

Hence, the pipeline loss lowers the $Z_g$ of a real nonad (here $2.5\sigma$) even more than that of a triad; we compensated for this effect only in the loss-corrected cross sections of triads (\S\ref{subsec:Losses}).
Had we compensated for this loss also in the $Z$-scores, the nonad would rise, for a median mock, to $Z_l\simeq4.3$ and $Z_g\simeq3.7$ (Appendix~\ref{app:Mock}), while the a priori $72.1\GeV$ triad would rise from $3.0\sigma$ to $\sim3.7\sigma$ (before combining with $L_{43}$).

An alternative, $\{m_1,m_2\}c^2\simeq\{67.4,75.6\}\GeV$ nonad is nearly as strong as the leading nonad ($Z_l\simeq3.16$, against $3.19$; $2.4\sigma$) and ranks first under equal-$Z$ weighting and in $\delta I$.
In this nonad, the $\sim72\GeV$ triad is the cross-annihilation term, and the $\sim23\GeV$ line is the $\gamma h$ channel of the $75.6\GeV$ species.
In either case, the $\gamma Z$ reading of $L_{43}$ fixes the mean mass $\overline{m}_\chi c^2\simeq 72.1\GeV$ of two (either self or cross)-annihilating WIMPs.

With one mass fixed at $72.1\GeV$, a scan over the other nonad mass reaches $2.9\sigma$, comparable to the nominal triad despite the free mass, and $5.0\sigma$ ($3.8\sigma$) when combined with $L_{43}$ ($L_{43b}$).
Fixing $72.1\GeV$ as the cross-annihilation mass
gives $3.0\sigma$ ($5.1\sigma$ with $L_{43}$, $3.9\sigma$ with $L_{43b}$), against $2.9\sigma$ as a self-annihilation mass, so the data do not distinguish the two roles (\S\ref{sec:43GeV}).

The $72.1\GeV$ mass may be a cross-annihilation mass also if the $67.5$ and $71.9\GeV$ triads
are redshifted as fully as their $\gamma\gamma$ lines, as would be the case for $\gamma\gamma$-dominated triads (\S\ref{sec:43GeV}).
For emission at $z\simeq0.03$, where the anticipated signal peaks (Appendix~\ref{app:PeakZ}), these leading $67.5$ and $71.9\GeV$ triads may then correspond to $\{m_1,m_2\}c^2\simeq\{69.5,74.1\}\GeV$ WIMPs, whose $\sim71.8\GeV$ cross-annihilation triad matches the $\gamma Z$ reading of $L_{43}$ (see \S\ref{sec:43GeV}).

\subsection{Supporting evidence}
\label{subsec:Evidence}

Several additional properties of the signal support an astrophysical WIMP origin.
Notably, the spectral features appear only in the X-ray-correlated spectrum $\delta I_\theta$.
The ambient spectrum $\delta I_0$, although $\sim2.4$ times less noisy, shows none of these features (Fig.~\ref{fig:Spectra}): $|Z_l|\le1.1$ at the energies of all reported features, including $Z_l\simeq-0.85$ for the nominal triad, and no $Z_l>2$ triad or nonad. Any instrumental, event-reconstruction, sky, or analysis artifact acting on \gama-rays independently of the X-ray contrast would appear brighter in $\delta I_0$ by a factor $\sim25$--$130$, at $Z_l\simeq130$--$460$, so such artifacts account for $\lesssim1.3\%$ of any reported $Z_l$ (see \S\ref{subsec:Robustness}), and the reported features are securely attributed to X-ray-correlated emission.

All three channels of the nominal triad contribute: $\{Z_{\gamma\gamma},Z_{\gamma Z},Z_{\gamma h}\}\simeq\{2.0,1.7,1.4\}$ (see \S\ref{sec:43GeV}).
The blind line scan independently picks up its $\gamma Z$ channel, as the $\sim43\GeV$ line, and the strongest line, at $\sim69\GeV$, where the $\gamma\gamma$ channels of the two leading triads blend (\S\ref{subsec:Blind}).
The wide, $17.7$--$72.1\GeV$ dynamic range of the kinematically fixed triad acts as a built-in veto against localized instrumental or background artifacts.

The spectral features are robust across the sky.
In particular, the nominal triad is distributed across the sky, appearing both north and south of the Galactic plane and in both longitude halves (\S\ref{subsec:Robustness}).
Masking Virgo and Fornax, or the Magellanic Clouds region, changes its significance only as expected from the reduced statistics (\S\ref{subsec:Robustness} and \S\ref{sec:43GeV}).

The $\sim72\GeV$ triad strengthens under positive (cosmological) but not negative redshift broadening, consistent with the expected LSS signal (\S\ref{subsec:Dz}); artifacts or local sky features would not typically show such behavior.
The preferred broadening, $\Delta z\simeq0.05$, corresponds to a mean redshift close to that at which the signal anticipated from LSS peaks ($\sim0.03$; \S\ref{subsec:Dz} and Appendix \ref{app:PeakZ}).
Mock triads spread over $\mathcal{Z}(z)$ verify the sensitivity of this test to cosmological broadening, whereas $z=0$ mock triads strengthen as much, with a $\Delta z\gtrsim0.05$ peak, in only $\sim3\%$ of cases (\S\ref{subsec:Dz}).

The nominal triad also strengthens in calibrated significance when each pixel is weighted by the X-ray contrast $\boldsymbol{X}_\theta$ smoothed with the LAT PSF of each line, rather than by $\boldsymbol{X}_\theta$ itself, as expected if the line photons trace the X-ray gas but are spread by the PSF into neighboring pixels. Under this weighting, the triad keeps $Z_l\simeq2.9$, whereas a noise fluctuation of the same nominal $Z_l\simeq3.0$ would typically drop to $Z_l\simeq2.1$,
so the calibrated significance of the triad rises by $\sim0.15\sigma$ (in a Gaussian approximation to the tails of $300$ shuffled-contrast skies; only $7\%$ of these skies exceed their own noise expectation under this weighting by as much as the triad does).
Conservatively, this PSF-smoothed weighting is not adopted as nominal, as it requires a different X-ray template for each line energy, whereas the nominal $\boldsymbol{X}_\theta$ defines a single correlated spectrum, shared by all scans and calibrations.

A simple, model-independent test, which uses no matched filter, points to emission lines.
Lines add positive fluctuations to the detrended spectrum, whereas the noise, as calibrated by the shuffled-contrast skies, is symmetric.
The $\delta I_\theta$ spectrum shows an excess of positive over negative $|y_k/\sigma_k|>2$ residuals, as expected from emission lines, at $2.6\sigma$ ($2.2\sigma$ after a \smash{\v{S}id\'{a}k} correction; \S\ref{subsec:Blind}), arising largely in the triad channels, as expected for an annihilation signal.
Simply summing, without matched filters, the pulls $y_k/\sigma_k$ of the bins holding the three channel energies of the nominal triad gives $3.1\sigma$ ($2.9\sigma$) at the medium (low) resolution of Fig.~\ref{fig:Spectra}, as calibrated by shuffled-contrast skies (\S\ref{sec:43GeV}).

The results are insensitive to analysis details: the detrending scale, modest rebinning, the eROSITA band and NXB treatment, the ROI cuts, and the LAT epoch (\S\ref{subsec:Robustness}).
As a consistency check, the LSS-enhanced sum $\delta I$, which adds the featureless ambient emission, recovers the same features (Appendix \ref{app:dI}).
The shuffle and Gaussian calibrations agree (Appendix \ref{app:MC}).
Mock signals show that the detrending lowers the measured $Z_l$ of a real triad by $\sim10$--$30\%$ (Appendix \ref{app:Mock}), so the quoted significance values understate that of a real signal.

\subsection{Implications}
\label{subsec:Implications}

The $\gamma Z$ reading of $L_{43}$, together with the present $\sim72\GeV$ triad, predicts additional $\gamma\gamma$ and $\gamma h$ lines at $72.1$ and $17.7\GeV$ in the rest frames of galaxy clusters.
This prediction pertains in particular to the three clusters reported to show $L_{43}$, but such a test is limited by the small number of line photons available there, as well as additional limitations such as Galactic foreground and point-source contamination.

The present results predict additional lines, which could be picked up in individual or stacked clusters across the sky, within or outside the present western-hemisphere region of interest.
In particular, besides a $\gamma\gamma$ line at the triad mass, the $67.5\GeV$ second triad predicts $\gamma Z$ and $\gamma h$ lines around $36.8$ and $9.5\GeV$, and the $69.2\GeV$ cross-annihilation triad of the leading nonad (\S\ref{subsec:Blind}) predicts such lines around $39.1$ and $12.5\GeV$.
The alternative nonad would add a $75.6\GeV$ triad, with $\gamma Z$ and $\gamma h$ lines around $48.1$ and $23.8\GeV$.
For the $\sim0.9\%$ mass offset (of mocks with the channel brightness ratios measured for the two leading triads),
these energies are underestimated by $\sim0.9\%$ in the $\sim70\GeV$ range, rising to
$\sim12\%$ in the
$\sim10\GeV$ range.

For the nominal triad, whether it arises from self- or cross-annihilation, the loss-corrected intrinsic channel cross sections are $\overline{\sigma v}_{p,i}/(10^{-19}D\cm^3\se^{-1})=10.8\pm8.7$, $28\pm29$, and $46\pm34$ for the $\gamma\gamma$, $\gamma Z$, and $\gamma h$ channels, respectively, assuming $\mathcal{Z}(z)$ (see \S\ref{sec:43GeV}).
Here, the WIMPs are assumed to perfectly trace the baryons, with substructure-correction factors (which are not substantial for volume-integrated $p$-wave annihilation) and dark-sector details all collected in the ignorance factor $D\propto 1/B$; for cross-annihilation, $D$ should also absorb the abundances of the two species.

For $D\sim1$, such values are comparable to published $p$-wave upper limits that pertain to the continuum, whose cross section for canonical WIMPs exceeds that of loop-suppressed lines by $2$--$3$ orders of magnitude.
The inferred line cross sections would thus favor frameworks with enhanced monochromatic branching or suppressed continuum production, discussed below.
The hadronic decays of the $Z$ and $h$ bosons of the $\gamma Z$ and $\gamma h$ channels should\cite{PDG2024} produce $\sim1\%$ of the observed X-ray-correlated continuum at $43\GeV$, $\sim15\%$ at $18\GeV$, and about half at $5\GeV$ (for our best-fit cross sections), independent of $D$; such a continuum is consistent with the data and cannot be easily measured.

The measured cross sections (Table~\ref{tab:Results}) follow from Eq.~\eqref{eq:CrossSection}, so they underestimate $\overline{\sigma v}_{p,i}/D$.
Correcting for pipeline and PSF losses (\S\ref{sec:43GeV}), and for $\mathcal{F}_b<1$ (Appendix~\ref{app:LossFactors}), only raises them.
The normalization $D=1$ also sets $\Omega_\chi=\Omega_c$ in Eq.~\eqref{eq:W}, \ie assumes a single Majorana species that makes up all the DM.
With the other factors of $D$ at their normalized values, the cross sections measured for the self-annihilation triad of a Majorana species of density parameter $\Omega_\chi$ constrain the product $(\Omega_\chi/\Omega_c)^2\,\overline{\sigma v}_{p,i}$.

For the cross-annihilation triad of two distinguishable Majorana species of density parameters $\Omega_1$ and $\Omega_2$, the factor $(\Omega_c/\Omega_\chi)^2$ is replaced by $\Omega_c^2/(2\Omega_1\Omega_2)$, and the measured values constrain the product $2\Omega_1\Omega_2\,\overline{\sigma v}_{p,i}/\Omega_c^2$.
As $\Omega_1+\Omega_2\le\Omega_c$, the intrinsic cross sections then exceed those inferred for a single species by a factor of at least $2$ for a cross-annihilation triad, and at least $4$ for the self-annihilation triad of the less abundant species.
Moreover,
the relevant $\Omega_\chi$, $\Omega_1$, and $\Omega_2$ may be much smaller than $\Omega_c$ if additional species exist, rendering the intrinsic cross sections correspondingly larger.
The aforementioned published limits, and the following GC estimate (\S\ref{subsec:GC}) and Galactic-halo bound (\S\ref{subsec:Halo}), derived for simplicity for a single species that makes up all the DM, scale similarly when applied to one of several species, so the comparisons hold quite generally.

The channel cross sections show a crude $\gamma\gamma<\gamma Z<\gamma h$ hierarchy, both in the measured values of the two leading triads (\S\ref{subsec:Blind}) and in the loss-corrected values above, albeit with large uncertainties.
This hierarchy is unlikely to reflect the instrumental efficiency: the effective area and energy resolution depend weakly on energy in the $10$--$70\GeV$ range, and the PSF, $\sim1.5$ times broader ($68\%$ containment) at $10$ than at $70\GeV$, dilutes the lower-energy lines more, opposing the hierarchy.

For Majorana WIMPs, $p$-wave annihilation proceeds from $^3P_J$ states, of which the $J=1$ state cannot annihilate into $\gamma\gamma$, so some suppression of the $\gamma\gamma$ channel may be expected.
In this limited sense, finding a $\chi\chi\to\gamma\gamma$ cross section smaller than its massive-boson counterparts, despite its larger phase space, is retroactively consistent with the present focus on $p$-wave annihilation.
The crude $\gamma\gamma<\gamma Z<\gamma h$ hierarchy could arise, for example, from a predominantly charge-conjugation-odd amplitude (possible for Dirac WIMPs), coupled mainly to heavy vector-like fermions.

Additional LAT data, a high-resolution eROSITA map of the eastern Galactic hemisphere, and future \gama-ray missions with improved energy resolution \cite{FanEtAl26} could confirm or refute our results.
Even with present data, the signals can be substantially improved, for example by relaxing some of the aggressive cuts, removing contaminations such as the NXB, or utilizing additional LSS tracers.
The spectral structure of the triad is rigidly fixed by kinematics, so a sufficiently careful analysis might be used in the future, for example to calibrate the overall energy normalization of the LAT or its successors.

\subsection{The Galactic halo}
\label{subsec:Halo}

The ambient spectrum $\delta I_0$ (Fig.~\ref{fig:Spectra}) further constrains $D$, as it includes the smooth Galactic halo, whose $p$-wave annihilation would produce the same triad.
Integrating the halo \gama-ray emissivity along the line of sight gives halo line $i$ the integrated brightness
\begin{equation}\label{eq:HaloLines}
  \mathcal{I}_{0,i} \simeq \frac{Y_i\epsilon_i\,\beta_0^2\,\overline{\sigma v}_{p,i}}{8\pi\eta_\chi m_\chi^2}\left\llangle\int\rho_\chi^2\,d\ell\right\rrangle \, ,
\end{equation}
where $\beta_0$ is the dimensionless line-of-sight velocity dispersion in the Galactic halo and $\overline{\sigma v}_{p,i}$ is the loss-corrected cross section, inferred from Eq.~\eqref{eq:CrossSectionLosses} for a given $D$.
For DM composed of a single Majorana species, in an NFW halo of local mass-energy density $0.4\GeV\cm^{-3}$ and $150$ km s$^{-1}$ line-of-sight velocity dispersion, $\mathcal{I}_{0,i}\simeq15D\,\mathcal{I}_i/(\mathcal{F}_i\mathcal{F}_{\mbox{\tiny PSF},i})$.

For large $D$, the halo triad would be strong in $\delta I_0$, which is $\sim2.4$ times less noisy than $\delta I_\theta$, and the pipeline would retain $\sim0.5$ of its brightness (in mock triads injected at $Z_l\simeq2$--$30$, nearly independently of strength).
The nominal triad ($Z_l\simeq3.0$ in $\delta I_\theta$, assumed spread over $\mathcal{Z}(z)$ LSS) would thus present in $\delta I_0$ at $Z_l\simeq3.6\times10^2D$.
Finding only $Z_l\simeq-0.85$ in $\delta I_0$ thus constrains $D\lesssim0.03$ for a WIMP annihilation triad (of a single Majorana-species DM), even allowing for systematic uncertainties (see below), or $D\lesssim0.15$ for the lower limit on $\Sigma_0$ set by detrending losses alone (\S\ref{sec:43GeV}).
For simplicity, the above assumes $\mathcal{F}_b=1$; Appendix~\ref{app:LossFactors} tightens it by a factor $\gtrsim1.6$.

The cluster signal must then exceed its $D=1$ estimate (\S\ref{subsec:SigmaCorrelations}) by a factor $\gtrsim30$ ($\gtrsim6$ for the lower limit on $\Sigma_0$).
Such an enhancement may be explained as a boost factor $B\gtrsim30$, indicating DM more concentrated in cluster cores than the X-ray-emitting gas.
An alternative explanation is a velocity dependence steeper than $\beta^2$, such as expected in a Breit--Wigner resonance just above threshold \cite{GriestSeckel91, IbeEtAl09}, which would also suppress the signal in the inner Galaxy and in dSphs, and predict a cluster signal rising faster than $\beta^2$ with cluster mass.

The aforementioned bounds apply to $D$ for a single Majorana species.
For several WIMP species sharing the DM (\S\ref{subsec:Implications}), they pertain to $(\Omega_\chi/\Omega_c)^2D$ for the self-annihilation triad of a species of density parameter $\Omega_\chi$, and to $2\Omega_1\Omega_2D/\Omega_c^2$ for a cross-annihilation triad.
While $D$ is thus larger for species of low abundance, the scaling is identical for the cluster and halo signals, so the boost $B$ required above to explain the effect is unchanged.

Formally, at $95\%$ confidence, the absence of the halo triad in $\delta I_0$ limits $D\lesssim5\times10^{-3}$ for a standard NFW or Einasto\cite{Einasto65} halo, and $D\lesssim1.7\times10^{-2}$ across plausible halo parameters (local mass-energy density $0.3$--$0.5\GeV\cm^{-3}$, vertical-to-planar halo axis ratio $0.7$--$1.3$, line-of-sight velocity dispersion $120$--$180$ km s$^{-1}$).
These limits reflect the expected sensitivity;
the observed $Z_l\simeq-0.85$
of $\delta I_0$ would strengthen them by a factor $\sim2$, whereas local DM structure, halo asymmetry, or residual spectral artifacts could relax them to the $D\lesssim0.03$ bound above.

Cross-annihilation of non-scalar WIMPs generally also allows the heavier species to decay as $\chi_2\to\chi_1\gamma$, producing a line at $\epsilon=(m_2^2-m_1^2)c^2/(2m_2)\simeq5.3\GeV$ for the leading nonad and $\sim6.4\GeV$ for the fixed-mass nonad. The ambient spectrum $\delta I_0$, which includes the Galactic halo, shows no line at these energies ($|Z_l|\le0.06$), limiting the lifetime of the heavier species to $\tau_2\gtrsim (\Omega_2/\Omega_c)10^{11}$ Hubble times ($95\%$ confidence, for the above NFW halo). A model of two cross-annihilating species must therefore strongly suppress this dipole transition.

\subsection{The Galactic center excess}
\label{subsec:GC}

The GeV excess attributed \cite{AjelloEtAl16GC, AlemannoEtAl26} to the GC was argued to originate from unresolved astrophysical sources, such as millisecond pulsars or stellar bulge populations, rather than DM annihilation \cite{BartelsEtAl16, LeeEtAl16, MaciasEtAl18}.
Note, however, that a DM interpretation of the GC signal indicates a $49\pm 6\GeV$ WIMP mass \cite{CaloreEtAl15} ($49^{+16}_{-12}\GeV$ in DAMPE \cite{AlemannoEtAl26}), which broadens to $50\text{--}190\GeV$ when accounting for interstellar-emission modeling systematics \cite{KarwinEtAl17},
and a continuum $p$-wave cross section $\sim 10^{-19} \text{ cm}^3 \text{ s}^{-1}$ (for $\beta^2\sim10^{-7}$); both estimates are broadly consistent with our present results, for $D\sim1$.

The continuum cross section is typically $2$--$3$ orders of magnitude stronger than loop-suppressed triad lines for canonical WIMPs.
For $D=1$, the GC and triad results thus match only if the GC has a DM core \cite{PontzenGovernato12, NestiSalucci13} rather than a cusp; a core diminishes the localized central signal much more dramatically than its subtle effect on the volume-integrated signal estimated in \S\ref{subsec:SigmaCorrelations}.

However, the results remain consistent even for a DM cusp, in frameworks featuring either enhanced branching fractions to monochromatic channels, such as in inert doublet models \cite{GustafssonEtAl07} or Rayleigh DM interacting via effective operators \cite{WeinerYavin12}, or a fundamental suppression of broad continuum production, characteristic of leptophilic \cite{FoxPoppitz09} or secluded \cite{PospelovEtAl08} DM frameworks.
Even in such frameworks, the continuum from the decays of the triad's $Z$ and $h$ bosons (\S\ref{subsec:Implications}) would alone exceed the GC excess in a cusp for $D=1$, imposing a $D\lesssim0.03$ upper limit.

For $D\lesssim10^{-2}$, within the halo limits of \S\ref{subsec:Halo}, the intrinsic line cross sections are correspondingly smaller.
The GC continuum then exceeds them by a factor of $\sim2$--$9$
for $D\sim10^{-2}$, somewhat less than typical for loop-suppressed lines, and by $\sim1.3$--$2.0$ orders of magnitude, approaching the typical $2$--$3$, for $D\sim10^{-3}$, with no need for a core.

Similarly, the difficulty \cite{MaciasEtAl18} of reproducing the bulge-like morphology of the GeV excess by annihilation in a DM cusp is eased for $p$-wave annihilation, whose central emission is broadened by the suppressed DM velocity dispersion in the baryon-dominated potential \cite{BoddyEtAl18}.
Thus, the DM interpretation of the GC excess appears to be consistent with the present study in terms of WIMP mass and cross section, and plausibly morphology, both for a cored profile (for a canonical WIMP) and for a cusp (for continuum and line cross sections balanced by any of the above mechanisms).

\acknowledgements

I am grateful to Kfir Blum, Tomer Volansky, Avi Loeb, and Ely Kovetz for helpful discussions.
This research received funding from ISF grant No. 2126/22.

This work is based on data from eROSITA, the soft X-ray instrument aboard SRG, a joint Russian-German science mission supported by the Russian Space Agency (Roskosmos), in the interests of the Russian Academy of Sciences represented by its Space Research Institute (IKI), and the Deutsches Zentrum f\"ur Luft- und Raumfahrt (DLR). The SRG spacecraft was built by Lavochkin Association (NPOL) and its subcontractors, and is operated by NPOL with support from the Max Planck Institute for Extraterrestrial Physics (MPE). The development and construction of the eROSITA X-ray instrument was led by MPE, with contributions from the Dr.~Karl Remeis Observatory Bamberg \& ECAP (FAU Erlangen-Nuernberg), the University of Hamburg Observatory, the Leibniz Institute for Astrophysics Potsdam (AIP), and the Institute for Astronomy and Astrophysics of the University of T\"ubingen, with the support of DLR and the Max Planck Society. The Argelander Institute for Astronomy of the University of Bonn and the Ludwig Maximilians Universit\"at Munich also participated in the science preparation for eROSITA.

\bibliography{Virial}

\appendix

\section{eROSITA data reduction}
\label{app:eROSITA}

To construct western-hemisphere X-ray maps, eRASS1 data are projected onto an order 10 HEALPix grid. Two data reduction streams are used: merging the 48 pre-compiled, surface-brightness hierarchical progressive survey (HiPS) tiles (our nominal choice, retaining the NXB) or aggregating the $\sim 2400$ individual sky tiles. For the latter, the NXB can be removed by subtracting a uniform instrumental baseline rate derived from filter-wheel closed (FWC) calibration spectra. As our analyses remove constant baseline offsets, the explicit subtraction of the NXB has only a minor effect on our results.

To identify point sources, a background noise level is derived from the median absolute deviation (MAD) as a robust estimator, so a standard deviation equivalent $\sigma_b \simeq 1.483\,\mathrm{MAD}$ can be estimated.
Pixels exceeding a $5\sigma_b$ threshold above the ambient median, along with unobserved instrumental scanning gaps, are masked. Maps smoothly populating masked pixels via an iterative nearest-neighbor interpolation of valid pixels are prepared for visual purposes.

To translate photon rates to surface brightness, pixel values are scaled by the appropriate energy conversion factor (ECF) and divided by the HEALPix pixel solid angle. We adopt the fiducial background ECFs, which assume a representative spectral mixture of a $\sim 0.1\text{--}0.2\keV$ thermal plasma (local hot bubble and Galactic corona) and an extragalactic power-law background of photon index $\Gamma \simeq1.4$.

\section{Spectral detrending}
\label{app:GPR}

Removing broadband astronomical and instrumental continua to isolate narrow spectral features is mainly achieved using the non-parametric GPR
\cite{RasmussenWilliams06}.
Given the binned ambient brightness spectrum $U=\llangle \boldsymbol{I}\rrangle$, which is always positive, let $\mathcal{C}_k \equiv (s_0/\epsilon_k)U_k\delta\Omega\sum_p\mathcal{E}_p$ be the corresponding effective photon counts in bin $k$, define logarithmic variables $x_k = \ln \epsilon_k$ and $\tilde{y}_k \equiv \ln U_k$, and adopt a squared-exponential covariance kernel $K_{ij} \equiv \mathrm{Var}(\tilde{\mathbf{y}}) \exp[-(x_i-x_j)^2 / (2\lambda^2)]$. Here, the amplitude $\mathrm{Var}(\tilde{\mathbf{y}})$ is the variance of $\tilde{y}_k$ across bins, and the correlation scale $\lambda$ greatly exceeds the energy resolution, to preserve narrow spectral features. The smooth log-continuum is then
\begin{equation}
    \ln \mathrm{GPR}(U) = \bar{\tilde{y}} + \mathbf{K}[\mathbf{K} + \mathrm{diag}(\mathcal{C}^{-1})]^{-1} (\tilde{\mathbf{y}} - \bar{\tilde{y}})  \coma
\end{equation}
where $\mathrm{diag}(\mathcal{C}^{-1})$ is the diagonal Poisson noise covariance matrix of the log-counts, as $\mathrm{Var}(\ln\mathcal{C})\simeq\mathcal{C}^{-1}$, and $\bar{\tilde{y}}$ is the mean log-intensity.

The X-ray-correlated spectrum $I_\theta$ and the sum $\llangle \boldsymbol{I}\rrangle+I_\theta$ can be negative in individual bins, so they cannot be detrended logarithmically.
For such a spectrum $U'$, we instead apply GPR, with a squared-exponential kernel of the same correlation length $\lambda$, to its ratio $V_k\equiv U'_k/\mathrm{GPR}(U)_k$ to the smooth ambient continuum, without taking a logarithm.
The GPR of $V$ requires the noise variance of $U'$ in each bin: the variance propagated in \S\ref{subsec:Correlations} for $I_\theta$, that of Appendix~\ref{app:dI} for the sum, and the Poisson variance $U^2/\mathcal{C}$ for the ambient spectrum, all evaluated for the smooth continuum.
The kernel amplitude is chosen by maximizing the marginal likelihood.

The resulting continuum, $\mathrm{GPR}(U)\,\mathrm{GPR}(V)$, is then subtracted from $U'$.
The ambient spectrum is detrended in the same way, with $U'\to U$.
The nominal correlation length is $\lambda\simeq0.62$, \ie 30 native bins ($0.27$ decades in energy); varying it between $0.18$ and $0.63$ decades has a negligible effect on the results (\S\ref{subsec:Robustness}).
In addition to detrending the spectra using GPR, we also linearly detrend them in $x_k$ over $3$--$100\GeV$, weighting by $\sigma_k^{-2}$, to avoid spurious matched filter signals.

\section{The LSS-enhanced spectrum $\delta I$}
\label{app:dI}

In each energy bin $k$, let $\mathcal{C}_p$ be the photon counts in ROI pixel $p$, of expectation $(s_0/\epsilon_k)\delta\Omega\mathcal{E}_pI_p$, and $X_{\theta,p}$ the contrast of this pixel.
The ambient mean and the X-ray-correlated slope are both linear in the counts, and their sum is
\begin{equation}\label{eq:SumEstimator}
  \llangle \boldsymbol{I}\rrangle+I_\theta = \frac{\epsilon_k/s_0}{\delta\Omega\sum_p\mathcal{E}_p}\sum_p\left(1+\frac{X_{\theta,p}}{\llangle \boldsymbol{X}_\theta^2\rrangle}\right)\mathcal{C}_p \fin
\end{equation}
The Poisson variance of this sum is
\begin{equation}\label{eq:VarSum}
  \mathrm{Var}(\llangle \boldsymbol{I}\rrangle+I_\theta) \simeq \frac{\epsilon_k}{s_0\delta\Omega\sum_p\mathcal{E}_p} \left\llangle \left(1+\frac{\boldsymbol{X}_\theta}{\llangle \boldsymbol{X}_\theta^2\rrangle}\right)^2 \boldsymbol{I} \right\rrangle \coma
\end{equation}
which includes $\mathrm{Var}(\llangle \boldsymbol{I}\rrangle)$, $\mathrm{Var}(I_\theta)$, and their covariance.
With $\llangle \boldsymbol{X}_\theta^2\rrangle\simeq0.19$, $\mathrm{Var}(I_\theta)$ dominates, exceeding $\mathrm{Var}(\llangle \boldsymbol{I}\rrangle)$ by a factor $\sim5$.

These errors are validated empirically.
Over $3$--$100\GeV$, the normalized residuals of $\delta I$ have a standard deviation of $1.04$ (extreme values $-2.5$ and $+2.7$), compared to $\sim1.0$ for the shuffled-contrast skies, which contain no true correlation.
Incomplete errors that omit $\mathrm{Var}(I_\theta)$, namely the Poisson errors of the pseudo-counts $\sum_p\mathcal{C}_p+(s_0/\epsilon_k)\delta\Omega I_\theta\sum_p\mathcal{E}_p$ (used in \S\ref{subsec:Robustness}), instead give a standard deviation of $2.65$ in the data and $\sim2.6$ in the shuffled-contrast skies, underestimating the noise by the expected factor $(1+1/\llangle \boldsymbol{X}_\theta^2\rrangle)^{1/2}\simeq2.5$.

\begin{figure}[t!]
    \includegraphics[width=\columnwidth]{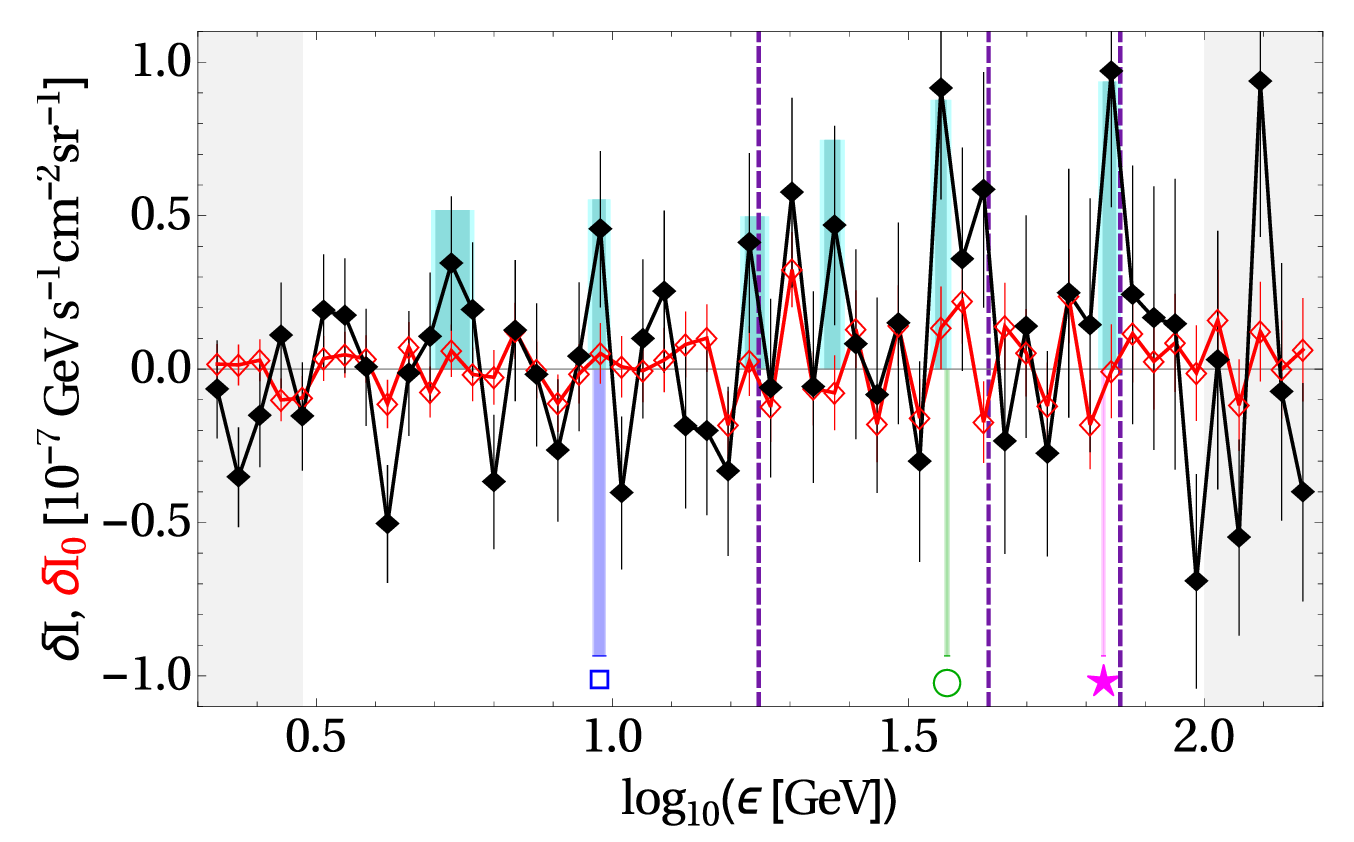}\\
    \includegraphics[width=\columnwidth]{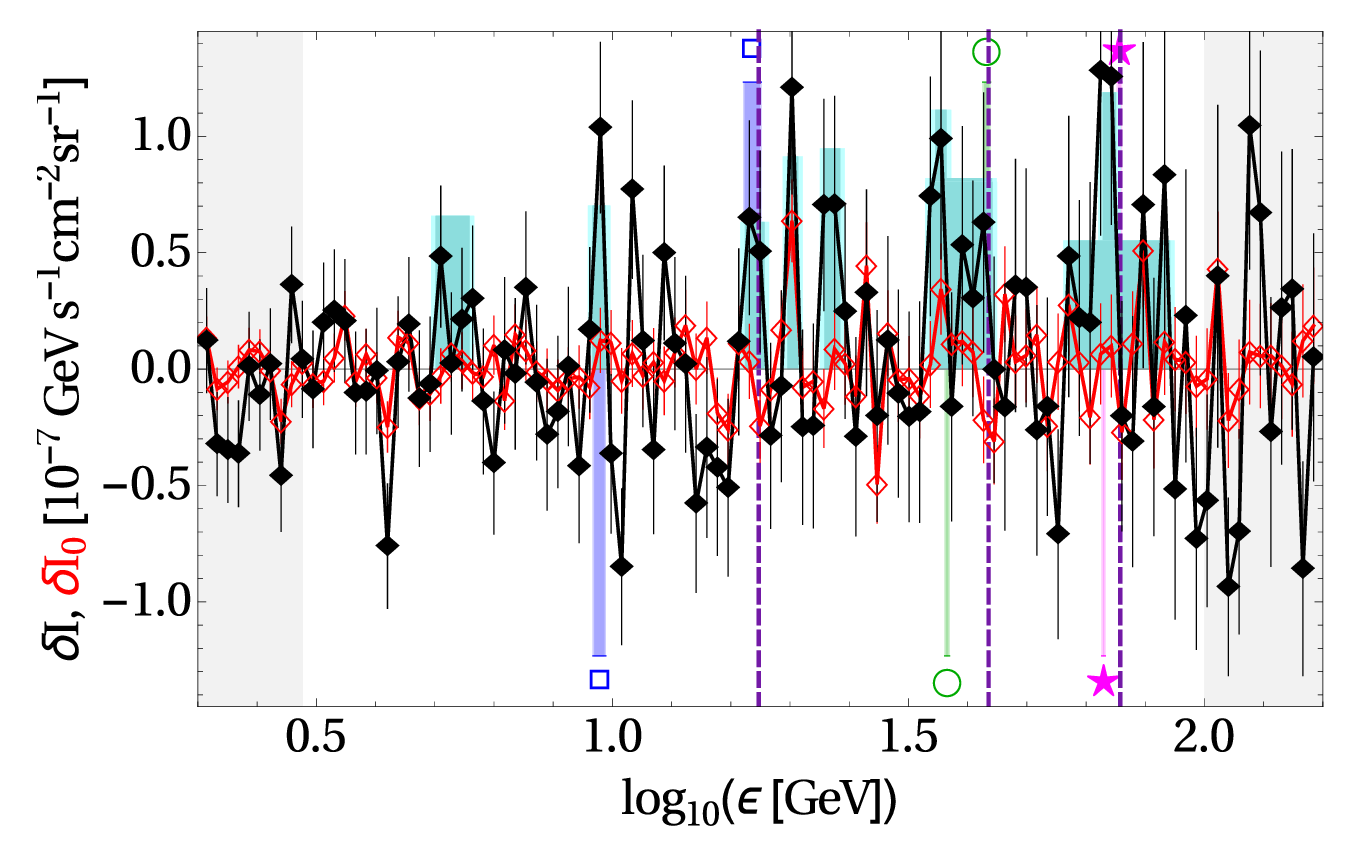}
	\caption{\label{fig:SpectradI}
        Same as Fig.~\ref{fig:Spectra}, but for the LSS-enhanced sum $\delta I$.
    }
\end{figure}

\begin{table*}[t]
    \centering
    \caption{Matched-filter features in the LSS-enhanced sum $\delta I$, selected as in Table~\ref{tab:Results}.\label{tab:dI}}
    \setlength{\tabcolsep}{9pt}
    \begin{tabular}{l @{\hspace{3em}} c @{\hspace{3em}} c @{\hspace{3em}} c @{\hspace{3em}} c @{\hspace{3em}} c}
        \toprule
        Feature & $\epsilon$ & $Z_l$ & $Z_g\,(Z_g^{\smash[t]{(a)}})\,[Z_g^{\smash[t]{(G)}}]$ & $\{\epsilon_{\gamma\gamma},\epsilon_{\gamma Z},\epsilon_{\gamma h}\}$ & $\{Z_{\gamma\gamma},Z_{\gamma Z},Z_{\gamma h}\}$ \\
         & (GeV) & & & (GeV) & \\
        \midrule
        \multirow{6}{*}{\shortstack[l]{Lines\\($f=0.1$)}} & $69.3_{-1.9}^{+1.6}$ & $2.4$ & $-0.1\,(1.2)\,[0.5]$ & & \\
         & $35.9_{-0.8}^{+0.8}$ & $2.3$ & & & \\
         & $23.5_{-0.7}^{+0.7}$ & $1.9$ & & & \\
         & $9.5_{-0.3}^{+0.2}$ & $1.4$ & & & \\
         & $5.2_{-0.2}^{+0.5}$ & $1.3$ & & & \\
         & $17.4_{-0.6}^{+0.7}$ & $1.3$ & & & \\
        \midrule
        \multirow{5}{*}{\shortstack[l]{Triads\\($f=0.015$)}} & $67.5_{-0.1}^{+0.1}$ & $3.3$ & $1.7\,(2.5)\,[2.2]$ & $\{67.5,\,36.8,\,9.5\}$ & $\{2.3,\,2.1,\,1.4\}$ \\
         & $71.8_{-0.3}^{+0.3}$ & $2.8$ & $0.6\,(1.7)\,[1.5]$ & $\{71.8,\,42.8,\,17.2\}$ & $\{1.9,\,1.6,\,1.3\}$ \\
         & $65.3_{-0.1}^{+0.4}$ & $2.2$ & $-0.6\,(0.9)\,[0.8]$ & $\{65.3,\,33.4,\,5.3\}$ & $\{1.6,\,0.9,\,1.3\}$ \\
         & $85.6_{-1.3}^{+2.7}$ & $2.1$ & $-0.7\,(0.8)\,[0.7]$ & $\{85.6,\,61.3,\,39.8\}$ & $\{0.8,\,1.2,\,1.7\}$ \\
         & $75.4_{-0.5}^{+0.6}$ & $2.1$ & $-0.7\,(0.8)\,[0.7]$ & $\{75.4,\,47.8,\,23.4\}$ & $\{1.1,\,0.7,\,1.9\}$ \\
        \midrule
        \multirow{2}{*}{\shortstack[l]{Nonad\\($f=0.05$)}} & $67.4_{-0.2}^{+0.5}$ & \multirow{2}{*}{$3.3$} & \multirow{2}{*}{$1.8\,(2.6)\,[2.4]$} & & \\
         & $75.6_{-0.5}^{+0.9}$ & & & & \\
        \midrule
        \multirow{2}{*}{\shortstack[l]{Triads at fixed\\masses (\S\ref{sec:43GeV})}} & $72.1$ & $2.6$ & $2.8\,[2.6]$ & $\{72.1,\,43.2,\,17.7\}$ & $\{1.8,\,1.5,\,1.2\}$ \\
         & $87.8$ & $2.1$ & --- & $\{87.8,\,64.2,\,43.2\}$ & --- \\
        \bottomrule
    \end{tabular}
    \TblNote{
            Same notations as in Table \ref{tab:Results}, with identical $n_G$ (to one decimal) expectations; five or more $Z_l>2$ triads arise in $1.2\%$ of the noise spectra ($\smash{Z_g^{(G)}}\simeq2.3$).
            Values of $Z_g$ are from $2\times10^4$ shuffled-contrast skies. The bottom rows are the triads at the externally fixed masses of the $\gamma Z$ ($72.1\GeV$) and $\gamma h$ ($87.8\GeV$) readings of $L_{43}$, with the calibrated significance of the $\gamma Z$ reading (for either reading: $2.7\sigma$ from shuffled-contrast skies and $2.4\sigma$ from the Gaussian calibration). Cross sections are not inferred from $\delta I$, which includes the X-ray-uncorrelated ambient emission.
            }
\end{table*}

The sum spectrum and its features are shown in Fig.~\ref{fig:SpectradI} and Table \ref{tab:dI}.
The most significant triad lies at $67.5_{-0.1}^{+0.1}\GeV$, at $Z_l\simeq3.3$, with global $Z_g\simeq1.7$ from $2\times10^4$ shuffled-contrast skies ($p_g\simeq0.04$), and $\smash{Z_g^{(G)}}\simeq2.2$ in the Gaussian Monte Carlo ($p_g\simeq0.016$).
The second triad lies at $71.8_{-0.3}^{+0.3}\GeV$ ($Z_l\simeq2.8$).
The three strongest lines, at $69.3$, $35.9$, and $23.5\GeV$, coincide, within the energy resolution, with the blended $\gamma\gamma$ channels of the leading triads, with the $\gamma Z$ channel ($36.8\GeV$) of the $67.5\GeV$ triad, and with the $\gamma h$ channel ($23.8\GeV$) of the $75.6\GeV$ mass of the nonad, respectively.
At the externally fixed masses of \S\ref{sec:43GeV}, the nominal triad reaches $Z_l\simeq2.6$, calibrated by shuffled-contrast skies as $p$-values of $2.6\times10^{-3}$ ($2.8\sigma$) for the $\gamma Z$ reading and $3.9\times10^{-3}$ ($2.7\sigma$) for either reading ($2.6\sigma$ and $2.4\sigma$ in the Gaussian calibration).
Its $\gamma\gamma$ and $\gamma h$ channels alone reach $2.2\sigma$ ($2.1\sigma$ in the Gaussian calibration).
These results are all consistent with those of $\delta I_\theta$, given the additional ambient noise.

\section{Averaged LAT IRF}
\label{app:IRF}

An inclusive, effective-area-weighted instrument response function (IRF) is derived from the official \texttt{P8R3\_ULTRACLEANVETO\_V2} calibration matrices (applicable also to \texttt{P8R3\_ULTRACLEANVETO\_V3}) as the probability density
\begin{equation}
\!\!\!\!\!\!\!P\left(s,\epsilon\right) = \frac{ \sum_{q=0}^{3} \sum_{t=1}^{8} A_{{\rm eff}, q}(\epsilon,\vartheta_t) S_{D,q}^{-1}(\epsilon,\vartheta_t) \, g(s/S_{D,q})}{ \sum_{q=0}^{3} \sum_{t=1}^{8} A_{{\rm eff}, q}(\epsilon,\vartheta_t) } \label{eq:P_inc}
\end{equation}
of the fractional energy error $s\equiv(\epsilon'-\epsilon)/\epsilon$.
Here, $q \in \{0, 1, 2, 3\}$ is the data quartile, $\vartheta_t$ is the binned photon incidence angle, $A_{\rm eff}$ is the fine-tabulated effective area, and
\begin{eqnarray}\label{eq:S_scale}
S_{D,q}(\epsilon, \vartheta) & = & c_{0,q} (\log_{10} \epsilon)^2 + c_{1,q} \cos^2\vartheta + c_{2,q} \log_{10} \epsilon \nonumber \\
& & + c_{3,q} \cos\vartheta + c_{4,q} \log_{10} \epsilon \cos\vartheta + c_{5,q}
\end{eqnarray}
is the scaling factor, with $\epsilon$ in MeV, where $c_{\ldots,q}$ are tabulated calibration coefficients.

The auxiliary functions
\begin{equation}\label{eq:P_mix}
g(\tilde{s}) \equiv F \, g_1(\tilde{s}) + (1 - F) \, g_2(\tilde{s})
\end{equation}
and
\begin{equation}
g_i(\tilde{s}) \equiv \frac{P_i}{S_i \, \Gamma(1/P_i)} \frac{K_i}{1+K_i^2} \exp\left[ -\left( \xi_i |\tilde{s} - B_i| \right)^{P_i} \right] \label{eq:g_fermi}
\end{equation}
depend on shape parameters tabulated on a coarse, 23-by-8-by-4 grid of $\{\epsilon,\vartheta,q\}$.
These empirically derived core ($i=1$) and tail ($i=2$) parameters are (keeping standard LAT collaboration variable names even if they override prior definitions in this text) $F$ (mixing fraction), $S_i$ (width), $K_i$ (skewness), $B_i$ (bias), and $P_i$ (exponential power index, controlling the kurtosis). The asymmetry multiplier $\xi_i$ sets the different left and right tails, $\xi_i = (K_i S_i)^{-1}$ for $\tilde{s} < B_i$, and $\xi_i = K_i / S_i$ for $\tilde{s} \ge B_i$.
The probability $P$, interpolated logarithmically in $\epsilon$, thus incorporates the heavy, bilateral IRF tails.

\section{Mock matched-filter tests}
\label{app:Mock}

To test the matched filters and quantify their performance, mock line or triad signals are injected into shuffled-contrast skies (Appendix \ref{app:MC}), preserving the empirical LAT energy binning, exposure, and photon statistics of the real data. Matched filters are then applied to the resulting mock $\delta I_\theta$ spectrum.

The signal is added to the correlated spectrum $I_\theta$ of each shuffled-contrast sky before detrending,
so the pipeline from detrending onward, including the scans, is tested with the errors propagated from each sky's photon counts.
As in the calibration (\S\ref{subsubsec:GlobalSignificance}), each sky first receives, before detrending, the continuum of the data.

Lines are injected with a prescribed local $Z$-score $Z_l$, distributed uniformly over $2$--$6$, at an energy $\epsilon_0$ distributed log-uniformly across $5$--$90\GeV$.
Triads are similarly injected, with either $Z_l$ or $\mathcal{I}$ distributed evenly among the three channels, at primary ($\gamma\gamma$ channel) $\epsilon_0$ distributed uniformly across the $66$--$98\GeV$ range.
Each case uses $600$ mocks; the statistics below pertain to those recovered, defined as mocks with one of their three strongest features (selected as in the data scans) lying within $10\%$ (lines) or $3\%$ (triads) of the injected $\epsilon_0$.

\begin{figure*}[t!]
    \centering
    \includegraphics[width=0.3568\textwidth]{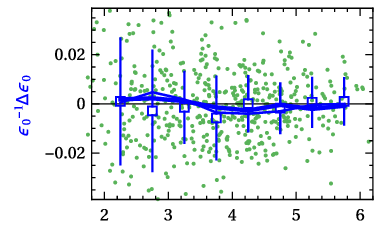}\hfill
    \includegraphics[width=0.3215\textwidth]{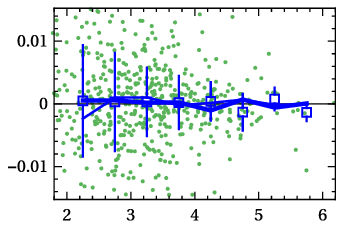}\hfill
    \includegraphics[width=0.3215\textwidth]{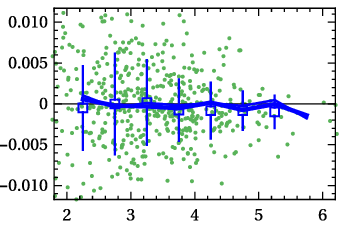}\\[2pt]
    \includegraphics[width=0.3568\textwidth]{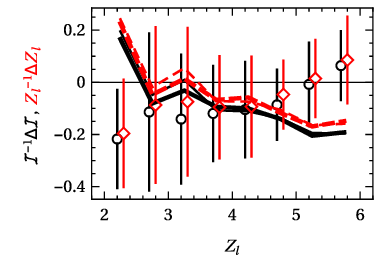}\hfill
    \includegraphics[width=0.3215\textwidth]{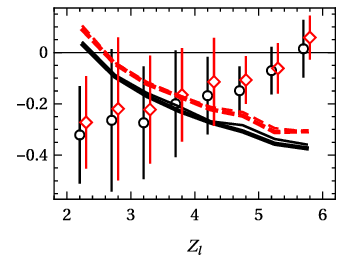}\hfill
    \includegraphics[width=0.3215\textwidth]{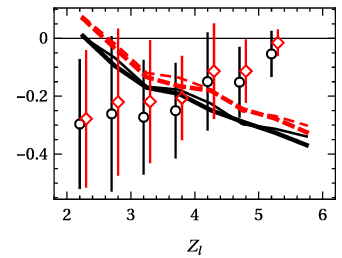}
	\caption{\label{fig:Mock}
        Mock matched-filter tests of line (left column), equal-$Z_l$ triad (middle), and equal-$\mathcal{I}$ triad (right) injection.
        Shown, as a function of both injected (curves) and inferred (symbols and error bars) $Z_l$, are the fractional errors $\epsilon_0^{-1}\Delta\epsilon_0$ in the inferred primary energy (equivalently in $\overline{m}_\chi$ for triads; top row, blue squares, and individual mocks as green points), $\mathcal{I}^{-1}\Delta\mathcal{I}$ in brightness (bottom row, black circles and solid curves), and $Z_l^{-1}\Delta Z_l$ (bottom row, red diamonds and dashed curves), where $\Delta$ denotes the inferred minus the injected value.
        Both standard (mean; thin curves) and robust (median and $1.483 \,\mathrm{MAD}$; thick curves and error bars) moments are shown (circles and diamonds offset horizontally by $\mp0.05$ for visibility).
    }
\end{figure*}

Figure \ref{fig:Mock} and the corresponding Table \ref{tab:Mock} show the resulting performance.
The table and the figure (error bars) report the robust statistics of the resulting mock sample (median and standard-deviation equivalent $1.483$ MAD); the figure also shows standard (mean) sample moments.
Primary energies are recovered without significant bias, with a dispersion of $\sim2\%$
for lines and $0.4$--$0.7\%$ for triads at measured $Z_l\simeq3$--$4$. Combining channels thus sharpens the mass determination.
The $\Delta$TS$=1$ bounds contain the injected $\epsilon_0$ in $68$--$77\%$ of the mocks, close to the nominal $68\%$.

The measured $Z_l$ of triads is typically lower than injected, by $\sim10$--$30\%$, and that of lines by $\lesssim20\%$.
Triad brightness is underestimated by $\sim27\%$ at $Z_l\simeq3$ and $\sim20\%$ at $Z_l\simeq4$, against $\sim20\%$ and $\sim17\%$ without the continuum of the data.
This loss is smaller than detrending alone would imply: in noise-free mocks (in which the signal is added to the smooth continuum of the data, without the photon noise of a shuffled-contrast sky),
detrending absorbs a median $\sim45\%$ of a triad's brightness (over $66$--$98\GeV$; $\sim35\%$ for the nominal triad).
Two effects offset much of this loss: the scan selects triads whose noise fluctuates upward, and the pipeline measures a positive mean brightness even without a signal (the null mean; Appendix~\ref{app:MC}).
The loss-corrected cross sections of \S\ref{sec:43GeV}, measured at fixed mass and thus without selection, account for the detrending and the null mean through $\mathcal{F}_i$ (see \S\ref{subsec:Losses} and Appendix~\ref{app:LossFactors}).

Mock nonads are injected, like lines and triads, into the correlated spectra of shuffled-contrast skies before detrending.
Three sets are used. The first has equal brightness in all nine lines, at the masses $67.5$ and $71.9\GeV$ of the two leading triads, with a nonad $Z_l=3$ evaluated before detrending.
In $47\%$ of cases, one of the three strongest nonads found has both masses within $3\%$ of the injected ones.

The contribution of the intermediate triad, at the recovered masses of each mock, is measured by its conditional $Z_c\equiv(Z_{\rm in}-\rho Z_{\rm out})/(1-\rho^2)^{1/2}$ score.
Here, $Z_{\rm in}$ ($Z_{\rm out}$) is the matched-filter $Z$-score, Eq.~\eqref{eq:MatchedFilter}, of the summed channel apertures of the intermediate triad (the two outer triads), and $\rho$ is their Pearson correlation coefficient for independent Gaussian noise of standard deviation $\sigma_k$ in each bin, so that $Z_c$ has unit variance in such noise.
In the recovered $Z_l=3$ mocks, $Z_c$ has a median $1.0$, lies between $0.2$ and $2.0$ in $68\%$ of cases, and is $\le0.4$ (\ie the observed value) in $26\%$.
Nonads injected with $Z_l=4$ ($Z_l=5$) are recovered in $67\%$ ($71\%$) of cases, and give a median $Z_c\simeq0.8$ ($Z_c\simeq0.9$), and $Z_c\le0.4$ in $30\%$ ($23\%$) of cases.

The second and third sets have equal brightness in all nine lines and an injected $Z_l$ distributed uniformly over $2$--$6$.
In the second set, each of $300$ shuffled-contrast skies is injected with a nonad at the masses $\{66.4,71.9\}\GeV$ of the leading nonad and, in a separate run, with a $71.9\GeV$ triad of the same injected $Z_l$.
Measured at the injected masses, the nonads lose $29\%$ of their $Z_l$ (median), against $16\%$ for the triads, and lose more than the triad of the same sky in $79\%$ of the skies.

Nonads of this set recovered with $Z_l\simeq2.5$--$3.5$ were injected with a median $Z_l$ higher by a factor $\simeq1.34$ ($1.0$--$1.9$, $68\%$); applied to the leading nonad, with the same number of trials, this would give $Z_l\simeq4.3$ and $Z_g\simeq3.7$.
Triads at $71.9\GeV$ of this set recovered with $Z_l\simeq2.5$--$3.5$ were injected with a median $Z_l$ higher by a factor $\simeq1.26$ ($1.0$--$1.7$), which would raise the a priori triad to $Z_l\simeq3.7$.

The third set comprises $400$ mocks with a single nonad each, of random masses with $m_2/m_1$ in the range $1.05$--$1.2$, and $400$ mocks with a single triad each, of random mass.
Measured at the injected masses, the nonads lose $49\%$ of their $Z_l$, against $24\%$ for the triads.
Only $\sim20$--$45\%$ of the nonads are recovered by the nonad scan (rising with $Z_l$), against $\sim90$--$100\%$ of the triads by the triad scan.
In noise-free mocks,
the detrending removes $\sim50\%$ of the brightness of a $\{66.4,71.9\}\GeV$ nonad ($\sim65\%$ at random masses), against $\sim35\%$ ($\sim40$--$45\%$) for triads.

Spreading the injection of line and triad mocks over $\mathcal{Z}(z)$ is discussed in Appendix \ref{app:PeakZ}.

\begin{table}[t]
    \centering
    \caption{Matched-filter mock-signal tests.\label{tab:Mock}}
    \begin{tabular}{l c r r r}
        \toprule
        Signal & $Z_l$ & $\epsilon_0^{-1}\Delta\epsilon_0$ & $\mathcal{I}^{-1}\Delta\mathcal{I}$ & $Z_l^{-1}\Delta Z_l$ \\
        (mock) & (measured) & (\%)$\,\,\,\,\,$ & (\%)$\,\,\,\,\,$ & (\%)$\quad$ \\
        \midrule
        Line & $\sim3$ & $-0.1\pm2.0$ & $-13\pm28$ & $-8\pm29$ \\
         & $\sim4$ & $-0.2\pm1.5$ & $-11\pm19$ & $-9\pm19$ \\
        \midrule
        Triad (equal-$Z_l$) & $\sim3$ & $0.0\pm0.7$ & $-27\pm25$ & $-22\pm25$ \\
         & $\sim4$ & $0.0\pm0.4$ & $-19\pm18$ & $-14\pm18$ \\
        \midrule
        Triad (equal-$\mathcal{I}$) & $\sim3$ & $0.0\pm0.6$ & $-27\pm22$ & $-22\pm23$ \\
         & $\sim4$ & $-0.1\pm0.4$ & $-22\pm17$ & $-18\pm16$ \\
        \bottomrule
    \end{tabular}
        \TblNote{
        Median and standard-deviation equivalent $1.483$ MAD of the mocks whose measured $Z_l$ lies in $2.5$--$3.5$ (rows $\sim3$) or $3.5$--$4.5$ (rows $\sim4$). Here, $\Delta$ denotes the inferred minus the injected value.
        }
\end{table}

\section{Global significance}
\label{app:MC}

Global significance is estimated in four ways (see \S\ref{subsubsec:GlobalSignificance}): nominally by shuffled-contrast skies ($Z_g$), by a Gaussian Monte Carlo ($\smash{Z_g^{(G)}}$), and analytically, by the \smash{\v{S}id\'{a}k} ($\smash{Z_g^{(a)}}$) and, for triads, up-crossing ($\smash{Z_g^{(n)}}$) approximations.
Here we compare these estimates and trace their differences.

For each shuffled-contrast sky, the full pipeline (correlation, error propagation, detrending, and scan) is rerun.
We use $2\times10^4$ shuffled-contrast skies for $\delta I_\theta$
and $\delta I$.
Shuffled-contrast skies also calibrate the fixed-mass tests of \S\ref{sec:43GeV}.
For the triad, which involves no scan, the null distribution for the $\gamma Z$ reading is that of the tested statistic (the triad $Z_l$, or the combined $Z_l$ of its two companion channels) at $72.1\GeV$, and when allowing either reading, that of the larger of its values at $72.1$ and $87.8\GeV$.

Each redshift-treated aperture of \S\ref{sec:43GeV} is calibrated against the null distribution of the same aperture in the shuffled-contrast skies.

For the nonad, the tested statistic is the largest interior local maximum of $Z_l$ over the free mass, with the other mass fixed at $72.1\GeV$, and when allowing either reading, the larger of these maxima for $72.1$ and $87.8\GeV$.
For the cross-annihilation role, the mean of the two masses is fixed at $72.1\GeV$, and the tested statistic is the largest interior local maximum over the lower mass, on the same $0.5\%$ grid.
Allowing either role, it is the larger of the self- and cross-annihilation maxima.

The four estimates of $Z_g$ are generally consistent.
For the leading $\delta I_\theta$ triad, they give $Z_g\simeq1.8$, $\smash{Z_g^{(G)}}\simeq1.8$, $\smash{Z_g^{(n)}}\simeq1.9$, and $\smash{Z_g^{(a)}}\simeq2.1$.
For the leading $\delta I_\theta$ nonad, the shuffles and the Monte Carlo give $Z_g\simeq2.5$ and $\smash{Z_g^{(G)}}\simeq2.2$.
At the fixed mass, both calibrations give $3.0\sigma$ for the $\gamma Z$ reading of the triad, and the shuffles (Monte Carlo) give $2.4\sigma$ ($2.4\sigma$) for its companion channels and $2.9\sigma$ ($2.6\sigma$) for the nonad.
Among leading triads and nonads, the largest differences between the shuffle and Monte Carlo calibrations
arise for the leading triad and nonad of $\delta I$
($Z_g\simeq1.7$ and $1.8$, against $\smash{Z_g^{(G)}}\simeq2.2$ and $2.4$; Appendix~\ref{app:dI}).

These differences reflect the shuffle null distribution of the matched-filter $Z_l$, which has a positive, $\sim0.2$--$0.5$ mean for single lines at $15$--$72\GeV$, and a standard deviation slightly below unity ($0.88$--$0.94$ for single lines, $0.7$--$0.8$ for the fixed-mass triads), whereas the Gaussian Monte Carlo assumes zero mean and unit variance.
The positive mean lowers the shuffle significance (as for the $\delta I$ triad), whereas the narrower spread raises it far in the tail (as for the fixed-mass nonad).
For the nominal triad, the positive mean arises from the final linear detrend (Appendix~\ref{app:GPR}), weighted by errors estimated from the same photon counts: with expected-count weights, its null mean drops from $0.54$ to $(-0.07)$.
The \smash{\v{S}id\'{a}k} estimates are somewhat higher than the Monte Carlo values for triads.

\section{Signal redshift structure}
\label{app:PeakZ}

To estimate the redshift structure of the cumulative annihilation signal and the redshift $z_{\text{peak}}$ of its peak contribution, we combine the integrated contributions of halos along the line of sight with the scarcity of massive halos in the small nearby volume. The analysis is performed for the present cross-correlation and, for completeness, for stacking analyses, which have slightly different redshift dependencies.
We adopt the ellipsoidal-collapse \cite{ShethTormen99, ShethMoTormen01} mass function, based on the variance of the BBKS
\citep{BardeenetAl86} linear power spectrum smoothed with a top-hat filter.

In general, the contribution of halos of mass $M$ at redshift $z$ is
\begin{equation}
    \mathcal{J}(M,z) \propto d_c^2\, \frac{d(d_c)}{dz}\, \mathcal{Q}\, \frac{dn_{\rm h}}{dM} \, ,
\end{equation}
where $d_c(z)$ is the comoving distance, $dn_{\rm h}/dM$ is the comoving halo mass function, and $\mathcal{Q}(M,z)$ is the contribution of a single halo, which depends on the analysis, as specified below.
Each halo contributes a photon flux
$\Phi\propto L_{\tilde{\epsilon}}/[(1+z)d_c^2]$, where $L_{\tilde{\epsilon}}(M,z)\equiv L_\epsilon(\tilde{\epsilon}) \propto \rho_{\rm cr}^{4/3} M^{5/3} \kappa^4 [\ln(1+\kappa)-\kappa/(1+\kappa)]^{-3}$
is the NFW-halo velocity-weighted $p$-wave specific luminosity, $\tilde{\epsilon}$ is the emitted photon energy, $\rho_{\rm cr}(z)$ is the critical density, and $\kappa(M,z)$ is the halo concentration parameter\cite{DuffyEtAl08}.

We approximate both the halo and the LAT point-spread function (PSF) as Gaussians of angular dispersions $\theta_{\rm h}$, taken as a quarter of the angular radius enclosing a mean density $500\rho_{\rm cr}$, and $\theta_{\text{PSF}}$, chosen so that its $68\%$ containment matches that of the LAT PSF ($\theta_{\text{PSF}}\simeq0\dgrdot07$ at $\sim70\GeV$).
These Gaussians enter only this halo model and the analytic estimate below; the PSF fractions $\mathcal{F}_{\mbox{\tiny PSF},i}$ use the actual LAT PSF, and all spectral filters and mocks use the actual IRF (Appendix~\ref{app:IRF}).

A cross-correlation analysis weights each halo by its soft X-ray surface brightness $X_{\rm h}\propto (1+z)^{-4}\rho_{\rm cr}^{5/3}M^{1/3}$ and by the overlap of its X-ray and PSF-smoothed \gama-ray profiles.
Here, $\mathcal{Q}\propto (2\theta_{\rm h}^2+\theta_{\text{PSF}}^2+2\sigma_{\rm pix}^2)^{-1}\theta_{\rm h}^2\, \Phi X_{\rm h}\, e^{-(4\theta_{\rm h}/\theta)^2}$, where the exponential factor approximates the effect of the $\theta=2\dgr$ filter $G$, and the pixelization of both maps is included by approximating each pixel as a square, and the square as a Gaussian of equal per-axis dispersion
$\sigma_{\rm pix}=(\delta\Omega/12)^{1/2}\simeq0\dgrdot017$ (a disk of the same area would give $0\dgrdot016$).
A background-limited stacking analysis is gauged instead by the squared signal-to-noise ratio, $\mathcal{Q}\propto (\theta_{\rm h}^2+\theta_{\text{PSF}}^2)^{-1}\Phi^2$.
Distant halos, unresolved by the PSF ($\theta_{\rm h}\ll\theta_{\text{PSF}}$), are thus suppressed in both correlation and stacking analyses.

\begin{figure}
    \includegraphics[width=\columnwidth]{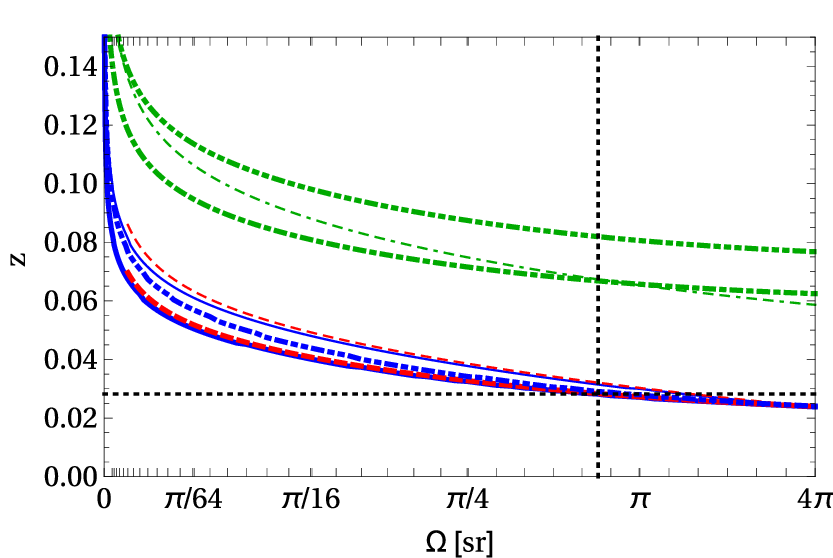}
	\caption{\label{fig:Zpeak}
        Redshift $z_{\textrm{peak}}$ of maximal contribution to the annihilation signal within solid angle $\Omega$ (minor ticks spaced by $0.1$ dex), for cross-correlation (thick curves) and stacking (thin curves) analyses, based on the $p$-wave luminosity of NFW halos in the ellipsoidal collapse model\cite{ShethTormen99} (solid blue) and on its low-redshift approximation (dashed red). Also shown are the median signal redshift (dot-dashed green), $z_{\textrm{peak}}$ and the median for the correlation analysis with the actual LAT PSF at $72.1\GeV$ in place of the Gaussian (thick double-dot-dashed blue and triple-dot-dashed green), and the solid angle and $z_{\textrm{peak}}$ of the present, eROSITA correlation analysis (dotted black vertical and horizontal lines).
    }
\end{figure}

A finite observed solid angle $\Omega$ harbors $\mathcal{N}(z)\simeq \Omega \int_0^z d_c^2 [d(d_c)/dz'](dn_{\rm h}/d\log M)\, dz'$ halos per mass decade out to redshift $z$.
The signal per unit redshift is then
\begin{equation}\label{eq:ZZz}
 \mathcal{Z}(z)=\int(1 - e^{-\mathcal{N}})\mathcal{J}\,dM \coma
\end{equation}
accounting for the Poisson statistics of rare massive clusters in the Milky-Way vicinity.
Its peak $z_{\text{peak}}$ is estimated from
\begin{equation}
    \frac{d\mathcal{Z}}{dz} \propto \int \left[ \left( 1 - e^{-\mathcal{N}} \right) \frac{\partial \mathcal{J}}{\partial z} + \mathcal{J} e^{-\mathcal{N}} \frac{\partial \mathcal{N}}{\partial z} \right] dM = 0 \fin
    \label{eq:ExactPeak}
\end{equation}
This estimate is shown in Fig.~\ref{fig:Zpeak} (solid blue curves); the figure pertains to both correlation (thick curves) and stacking (thin) analyses.

The results indicate $z_{\text{peak}}\simeq 0.03$ in both analyses, for the present parameters.
A low-redshift $\mathcal{N} \propto z^3 \Omega$ approximation is also shown (dashed red, for $\Omega\geq0.01\sr$).
The median redshift of the same distribution, below which half of the signal arises (dot-dashed green curves), is $\sim0.07$ in both analyses, for the present parameters.
Both $z_{\text{peak}}$ and the median rise as $\Omega$ decreases and nearby massive halos become rare, more so for stacking,
whereas the signals of both analyses decline as $\sim z^{-2}$ at high redshifts, where halos are unresolved.

Averaged over halos, the PSF retains the fraction $\mathcal{F}_{\mbox{\tiny PSF},i}$ of the correlation signal (\S\ref{subsec:Losses}), namely the ratio of $\int\mathcal{Z}\,dz$ to its $\theta_{\text{PSF}}\to0$ limit,
computed with the actual LAT PSF (of the \texttt{P8R3\_ULTRACLEANVETO\_V2} IRFs, averaged over event types as in Eq.~\eqref{eq:P_inc}) in place of the Gaussian in $\mathcal{Q}$.
The sharp core of the actual PSF retains more than a Gaussian of the same $68\%$ containment
($\mathcal{F}_{\mbox{\tiny PSF},i}\simeq0.52$ rather than $0.47$ for the $\gamma\gamma$ line of the nominal triad), mostly from distant halos of small angular size, so the correlation signal extends to higher redshifts (double-dot-dashed blue and triple-dot-dashed green curves in Fig.~\ref{fig:Zpeak}; median $\sim0.08$).

An annihilation line emitted with energy $\epsilon$ at redshift $z$ is observed at $\epsilon/(1+z)$, so a matched filter centered on $\epsilon$ recovers only a fraction $\psi(z)\le1$ of it.
Approximating the IRF by its Gaussian-equivalent width $\sigma_s\simeq0.037$
at $70\GeV$ gives $\psi(z)\simeq\exp[-z^2/(2\sigma_s)^2]$, and a recovered fraction $\sim0.4$ of the correlation signal ($0.3$--$0.6$ for $\theta_{\text{PSF}}\simeq0\dgrdot07$--$0\dgrdot15$ and halo dispersions $0.6$--$1.6$ times the nominal $\theta_{\rm h}$; $\sim0.35$ with the actual LAT PSF).

Mock triads spread over $\mathcal{Z}(z)$ and passed through the full pipeline (Appendix~\ref{app:Mock}) retain more of their brightness than this estimate, as the extended IRF wings still overlap the redshifted lines.
The unrecovered, redshifted part of the signal forms a low-energy tail, largely removed by the detrending. How much is retained depends on the measurement.

In the noise-free limit, mock triads located by a mass scan retain $\sim0.6$ of the brightness recovered for $z=0$ emission, or $\sim0.4$ of the injected brightness when the detrending losses of Appendix~\ref{app:Mock} are included.
Triads found by a blind scan at the observed $Z_l\simeq3$ retain more, $\sim0.8$ and $\sim0.7$, respectively, as the scan selects upward noise fluctuations and the measured brightness includes the positive null mean of Appendix~\ref{app:MC}. A fixed-mass measurement involves no such selection, and its templates remain centered on the unshifted energies, so at the measured strength it retains only $\mathcal{F}_i\simeq(0.4$--$0.5)\mathcal{F}_{0,i}$ (see \S\ref{subsec:Losses}). The loss-corrected fixed-mass cross sections are obtained by normalizing by these fractions as well as by $\mathcal{F}_{\mbox{\tiny PSF},i}$.

Such redshift spreading shifts the measured energies of single lines $\sim5$--$6\%$ below the emitted energy (\S\ref{subsec:Losses}). Triad masses shift by only $0.7$--$1.2\%$, in mocks injected with the channel brightness ratios measured for each of the two leading triads, in which the $\gamma\gamma$ channel is the brightest (\S\ref{subsec:Losses}).

\section{Loss factors}
\label{app:LossFactors}

\begin{table*}[t]
    \centering
    \caption{Loss factors of the nominal triad.\label{tab:LossFactors}}
    \setlength{\tabcolsep}{5pt}
    \begin{tabular}{@{}l p{0.52\textwidth} l c c c@{}}
        \toprule
        Factor & Definition & Basis & $\gamma\gamma$ & $\gamma Z$ & $\gamma h$ \\
        \midrule
        $\epsilon_i$ (GeV) & Channel energy &  & $72.1$ & $43.2$ & $17.7$ \\
        \midrule
        $\mathcal{F}_{0,i}$ & Pipeline fraction for $z=0$ emission & mocks & $0.77$ & $0.75$ & $0.82$ \\
        $\mathcal{F}_i$ & Pipeline fraction for emission spread over $\mathcal{Z}(z)$ & mocks & $0.30$ & $0.35$ & $0.34$ \\
        $\mathcal{F}_i$ ($z<0.1$) & Pipeline fraction for $\mathcal{Z}(z)$ truncated at $z=0.1$ & mocks & $0.53$ & $0.57$ & $0.55$ \\
        $\mathcal{F}_i$ (uniform) & Pipeline fraction for emission uniform over $z<0.05$ & mocks & $0.68$ & $0.69$ & $0.71$ \\
        $\mathcal{F}_{\mbox{\tiny PSF},i}$ & PSF fraction for emission spread over $\mathcal{Z}(z)$ & model & $0.52$ & $0.51$ & $0.46$ \\
        $\mathcal{F}_{\mbox{\tiny PSF},i}$ ($z<0.1$) & PSF fraction for $\mathcal{Z}(z)$ truncated at $z=0.1$ & model & $0.68$ & $0.67$ & $0.63$ \\
        $\mathcal{F}_{\mbox{\tiny PSF},i}$ (uniform) & PSF fraction for emission uniform over $z<0.05$ & model & $0.79$ & $0.78$ & $0.75$ \\
        \midrule
        $\mathcal{F}_{X,i}$ & PSF fraction for emission tracing the X-ray map & data & $0.30$ & $0.28$ & $0.22$ \\
        $\mathcal{F}_{\mbox{\tiny noise},i}$ & PSF fraction for pixel-scale X-ray noise & data & $0.25$ & $0.23$ & $0.16$ \\
        $\mathcal{F}_{\mbox{\tiny gas},i}$ & PSF fraction for gas-tracing X-ray emission, smoothed as $\mathcal{F}_{X,i}$ & model & $0.48$ & $0.47$ & $0.42$ \\
        $\mathcal{F}_b$ & Bremsstrahlung fraction of $\llangle \boldsymbol{X}_\theta^2\rrangle$ & data, model & $0.22$ & $0.22$ & $0.21$ \\
        $\mathcal{F}_b\mathcal{F}_{\mbox{\tiny PSF},i}$ & Upper limit, $1.1\,\mathcal{F}_{X,i}$ & data, model & $\lesssim0.33$ & $\lesssim0.31$ & $\lesssim0.24$ \\
        \bottomrule
    \end{tabular}
    \TblNote{Mocks: fixed-mass injections into the correlated spectra of shuffled-contrast skies, at the measured strength; model: the halo model of Appendix~\ref{app:PeakZ}; data: the X-ray map smoothed by the LAT PSF of each line.}
\end{table*}

Table~\ref{tab:LossFactors} lists the factors relating the measured channel brightness $\mathcal{I}_i$ of the nominal triad to its intrinsic cross section (Eqs.~\eqref{eq:Ratio} and \eqref{eq:CrossSectionLosses}).
The pipeline fractions $\mathcal{F}_{0,i}$ and $\mathcal{F}_i$ are measured by fixed-mass injections into shuffled-contrast skies, at the strength that reproduces the measured brightness, and $\mathcal{F}_{\mbox{\tiny PSF},i}$ follows from the halo model of Appendix~\ref{app:PeakZ}.

The pipeline fractions depend on the signal strength (\S\ref{subsec:Losses}), in part through the null mean, which amounts to $\sim10$--$25\%$ of the data's brightness in each channel of the nominal triad (Appendix~\ref{app:MC}).
We therefore inject triads, with equal channel $Z$-scores, at several strengths, and average the measured brightness of each channel over $200$ skies in each case.
Interpolating linearly between these strengths, we find for each channel the injected brightness whose mean measurement equals the real measured $\mathcal{I}_i$, and take $\mathcal{F}_i$ as the ratio of the two.
The corrected brightness $\mathcal{I}_i/\mathcal{F}_i$ is thus the injected brightness expected on average.
The $z=0$ fractions $\mathcal{F}_{0,i}$ (mean $0.78$) exceed the $\sim0.65$ that the nominal triad retains in noise-free mocks (Appendix~\ref{app:Mock}), mainly because $\mathcal{F}_{0,i}$ includes the null mean.

The signal photons omitted by the mocks (\S\ref{subsec:Losses}) carry Poisson noise that is already included in the count-based errors of the data, and largely averages out of $\mathcal{F}_i$ when averaged over $200$ skies.
These photons could still bias $\mathcal{F}_i$ through the count-based errors, which weight the detrending: concentrated in X-ray-bright pixels, they would raise the errors of the data at the channel energies, relative to neighboring energies, beyond those of the shuffled-contrast skies.
However, we find no significant excess, with $Z$-scores of $0.15$, $1.8$, and $0.62$ in the $\gamma\gamma$, $\gamma Z$, and $\gamma h$ channels.

The fraction $\mathcal{F}_i$ is estimated above at the nominal triad mass.
In noise-free mocks at random masses over $66$--$98\GeV$, the redshift spreading retains nearly the same fraction as for the nominal triad, but the detrending absorbs more (Appendix~\ref{app:Mock}), so these mocks retain a median $0.35$ of the injected brightness, against $0.41$ for the nominal triad.
At a typical mass in this range, $\mathcal{F}_i$ may thus be $\sim15\%$ lower.

Nonads are not loss-corrected, for two reasons: their aperture fits a common brightness to all nine lines (which need not be equally bright), and the $\gamma\gamma$ lines of close masses ($4\%$ apart for the leading nonad) overlap within the energy resolution, so the brightness of each of their triads
cannot be measured separately.
Nonad mocks (Appendix~\ref{app:Mock}) instead test the recovery and significance of nonads, which lose more of their brightness to the detrending than triads.

The fraction $\mathcal{F}_{X,i}$ (see \S\ref{subsec:Losses}) is the part of the covariance between the contrast and the X-ray map that survives smoothing the map with the LAT PSF of line $i$, into $\mathcal{P}_i(\boldsymbol{X})$,
\begin{equation}\label{eq:FX}
  \mathcal{F}_{X,i}\equiv\frac{\mathrm{Cov}_{\mathcal{E}}[\boldsymbol{X}_\theta,\mathcal{P}_i(\boldsymbol{X})]}{\mathrm{Cov}_{\mathcal{E}}(\boldsymbol{X}_\theta,\boldsymbol{X})}\coma
\end{equation}
with the exposure weights of the energy bin of line $i$, and the real contrast outside the ROI wherever eRASS1 covers the sky.
The fraction $\mathcal{F}_{\mbox{\tiny noise},i}$ is measured by shuffling $\boldsymbol{X}_\theta$ among pixels, and $\mathcal{F}_{\mbox{\tiny gas},i}$ is computed with the halo model for the same numerical smoothing (by FFT on $\sim220$ tangent-plane sky patches), which retains $\sim3\%$ less than exact smoothing.
As the noise term of $\mathcal{F}_{X,i}\simeq\mathcal{F}_b\mathcal{F}_{\mbox{\tiny gas},i}+(1-\mathcal{F}_b)\mathcal{F}_{\mbox{\tiny noise},i}$ is positive, $\mathcal{F}_b\mathcal{F}_{\mbox{\tiny gas},i}\lesssim\mathcal{F}_{X,i}$.
For line emission that traces the gas or is more concentrated, the halo model gives $\mathcal{F}_{\mbox{\tiny PSF},i}/\mathcal{F}_{\mbox{\tiny gas},i}\simeq1.09$--$1.10$, owing to the different halo weighting and the numerical smoothing, so $\mathcal{F}_b\mathcal{F}_{\mbox{\tiny PSF},i}\lesssim1.1\,\mathcal{F}_{X,i}$.

Equation~\eqref{eq:CrossSectionLosses} divides the measured brightness by $\mathcal{F}_b\mathcal{F}_{\mbox{\tiny PSF},i}$, taken as $\mathcal{F}_{\mbox{\tiny PSF},i}\simeq\{0.52,0.51,0.46\}$ with $\mathcal{F}_b=1$.
The bound $\mathcal{F}_b\mathcal{F}_{\mbox{\tiny PSF},i}\lesssim\{0.33,0.31,0.24\}$ shows that the actual product is smaller, by a factor $\gtrsim1.6$ ($\gtrsim1.9$ for the $\gamma h$ line).
The loss-corrected $\Sigma_0$ (Eq.~\eqref{eq:Sigma0}) and other $\overline{\sigma v}_{p,i}/D$ for $\mathcal{Z}(z)$ are therefore underestimated, and the bounds on $D$ (see \S\ref{subsec:Halo}) are too loose, by this factor.
The same factor applies to the alternative $\Sigma_0$ values quoted after Eq.~\eqref{eq:Sigma0}, as $\mathcal{F}_b\lesssim\mathcal{F}_{X,i}/\mathcal{F}_{\mbox{\tiny gas},i}\simeq0.6$ for any redshift distribution of the signal.
The estimate $\mathcal{F}_b\simeq0.22$ of \S\ref{subsec:Losses} implies a larger factor, $\sim4.5$ (or $4.2$--$4.8$ for halo dispersions $0.6$--$1.6$ times the nominal $\theta_{\rm h}$ of Appendix~\ref{app:PeakZ}).

The measured $\mathcal{F}_{X,i}$ is fairly robust: restoring the real contrast in the pixels masked as bright X-ray sources raises it by only $\sim0.004$, and the northern and southern halves of the ROI give $0.29$ and $0.31$ at $72.1\GeV$.

\section{Coarser native binning}
\label{app:Nominal}

Features nearly identical to those of our nominal analysis are recovered with a coarser native binning, at a somewhat lower significance. With a coarser $s_0\simeq3.7\%$ native binning (256 bins over $0.1$--$1000\GeV$), and an otherwise identical analysis, the features appear at the same energies and masses, but with $Z_l$ values lower by $\sim15$--$25\%$: the leading $\delta I_\theta$ triad at $71.8\GeV$ has $Z_l\simeq2.5$ ($\smash{Z_g^{(G)}}\simeq1.1$), the nominal triad has $Z_l\simeq2.3$ ($2.3\sigma$ for the $\gamma Z$ reading and $2.0\sigma$ for either reading in the Gaussian calibration), and the leading $\delta I$ triad at $67.5\GeV$ has $Z_l\simeq2.7$ ($\smash{Z_g^{(G)}}\simeq1.4$).
These differences are not due to the detrending scale, whose variation has a negligible effect at both resolutions.

\end{document}